\documentclass[%
 reprint,
 amsmath,amssymb,
 aps,
]{revtex4-2}

\usepackage{graphicx}% Include figure files
\usepackage{dcolumn}% Align table columns on decimal point
\usepackage{bm}% bold math
\usepackage[english]{babel}
\usepackage{natbib}
\graphicspath{ {images/} }
\setcitestyle{square,numbers,super}

\DeclareRobustCommand{\rchi}{{\mathpalette\irchi\relax}}
\newcommand{\irchi}[2]{\raisebox{\depth}{$#1\chi$}} % inner command, used by \rchi
\def\etab{\overline{\overline{\eta}}}

\begin{document}

\preprint{APS/123-QED}

\title{Thermodynamics of anharmonic lattices from first-principles}% Force line breaks with \\
%\thanks{A footnote to the article title,placeholder}%
\author{Keivan Esfarjani}%
\email{E-mail: ke4c@virginia.edu}
\affiliation{
 Department of Mechanical and Aerospace Engineering, 
 Department of Materials Science and Engineering, 
 Department of Physics, \\
 University of Virginia 
}
\author{Yuan Liang}
\affiliation{
 Department of Physics, University of Virginia
}

% \altaffiliation[Also at]{  Department of Materials Science and Engineering, 
% and Department of Physics, University of Virginia.}%Lines break automatically or can be forced with \\

%\affiliation{%
% Department of Mechanical and Aerospace Engineering, 
% Department of Materials Science and Engineering, 
% Department of Physics, \\
% University of Virginia
% \textbackslash\textbackslash
%}%

%\collaboration{placeholder}%\noaffiliation

\date{\today}% It is always \today, today,

\begin{abstract}
\textbf{Abstract:} Self-consistent phonon (SCP) theory and its application in computing thermodynamic properties of materials are reviewed from a historical perspective.  Various more recent implementations based on first-principles electronic structure methods using the density functional theory (DFT) have been discussed. The SCP equations can be derived either from a diagrammatic perturbation theory or a variational approach based on free-energy minimization. These methods can also be used to predict phase change due to phonon softening, and can be extended to study the coupling of phonons to other degrees of freedom in the system.
\end{abstract}

%Use showkeys class option if keyword

\maketitle
%\tableofcontents
\section{Introduction}

\subsection{Motivation}

Atomic vibrations play an important role in describing the thermodynamics of materials. In crystalline solids, we refer to these excitations as phonons. A quantum description based on phonons allows the calculation of thermodynamic properties such as free energy, heat capacity and entropy, as well as Raman intensities, all of which can be compared to experimental data on Neutron scattering and Raman spectra in order to better understand the underlying physics and properties of a material. Most commonly employed modern methods for phonon spectra calculation include harmonic and quasi-harmonic approximations, in which the potential energy is expanded up to second order in powers of atomic displacements (assumed to be small) about their equilibrium positions. The inclusion of only harmonic forces in the system implies the phonons have infinite lifetime. This limits the possibility to obtain properties such as thermal conductivity. Furthermore, at elevated temperatures (compared to the Debye temperature for instance), where atomic displacements become larger, anharmonic contributions to the forces are not negligible anymore and one needs to go beyond the harmonic approximation. Additionally, phonon frequencies vary with temperature and collisions between them limits their lifetime. These effects can be accounted for by treating anharmonic effects as a perturbation. The need for such theories come also in systems near a phase transition where atomic displacements are large and therefore anhamonicity more important. Another case of interest which originally motivated such theories was the study or rare gas crystals such as Neon or Argon (Helium was exceptionally challenging) where the quantum zero point motion of atoms is large and atoms probe anharmonic regions of the potential. The advantage of rare gas atom systems is that analytical pair interatomic potentials (typically of Lennard-Jones type) can accurately describe their properties, and therefore different theories can easily be tested. 

In this review, after going over the general lattice dynamics formalism, we will start by explaining the basic idea behind SCP in section \ref{idea}. This will be followed by a simple example to illustrate the method and its power in section \ref{example}. 
The historical development of SCP theory is detailed in section\ref{history} by briefly going in more details over the work of famous contributors including: Born, Hooton, Horton, Cowley, Choquard, Koehler, Werthammer, Gillis and many others. 
Next, modern implementations using DFT calculations will be described in Section \ref{modern-implementations}. This section will first deal with the model parameters extraction from DFT calculations in real or reciprocal spaces, and then its actual implementations will be discussed. We will be reviewing the merits and shortcomings of the two approaches. Finally, in Section \ref{our-extension} we propose an extension which includes internal and cell relaxations and  coupling to other order parameters (OPs). This method provides a theory of phase transitions incorporating the coupling of OPs with vibrational and strain degrees of freedom and also accounts for possible internal atomic relaxations and structural change at a low computational cost. 

%With previous methods discussed, we will also present an extension to the variational approach that also includes internal and lattice relaxations in the picture, thus enabling the prediction of phase transitions: the method also includes the coupling of atomic vibrations with displacive and other order parameters (OPs) such as ferromagnetic or orbital-ordering, which are usually not included in traditional first order phonon approximations. This allows to include the mutual effect of phonons and their renormalization on the OPs and their couplings. Thus the formalism can explain how phonons change order parameters and how the latter affect phonon frequencies and modes. This could potentially offer intuition on the causes of structural phase transitions in multifunctional materials. The initial input requires high-order force constants, the order parameter couplings and their mutual interactions with the atomic displacements, all of which can in principle be extracted from either density functional perturbation theory (DFPT) or using finite "displacements" in judiciously-prepared supercells. Due to the simplicity of this variational formulation, lattice anharmonicity and phase transitions can be easily studied at any given temperature with a high efficiency and relatively low computational cost, free of numerical noise present in Monte Carlo (MC) or molecular dynamics (MD) simulations. \\

%\subsection{Some History}

\subsection{Lattice dynamics theory and the self-consistent phonon idea} \label{idea}

In the Born-von-K\'arm\'an approximation, atomic nuclei possess a mass and are connected by massless electron "springs". The fundamental assumption from Born-von-K\'arm\'an's theory is that the oscillation of atomic nuclei is confined to a small region relative to the nuclear-nuclear separation. So to leading order, a set of harmonic equations of motion can be constructed to describe the system using the quadratic coefficients called the force constants. It is the basis of what is nowadays known as the theory of lattice dynamics, whose extensions will be discussed in this review. 
\begin{equation}
    H\approx H_{Harmonic}=\sum_{i,{\alpha}} \frac{P_{i\alpha}^2}{2m_i}	+  V_0 + \frac{1}{2}\sum_{i,j} \Phi_{i\alpha,j\beta} U_{i\alpha} U_{j\beta}
\end{equation}
In this equation, $U_{i\alpha}$ is the displacement of atom $i$ about its equilibrium position in the direction $\alpha$, $P_{i\alpha}$ is its conjugate momentum, and $\Phi_{i\alpha,j\beta}$ is called the harmonic force constant between atoms $i$ and $j$. Greek letters represent the cartesian components $x,y,z$ and they are summed over if repeated.
After decoupling the dynamical displacement variables by the appropriate unitary transformation $\epsilon_{i\alpha,\lambda}$, the Hamiltonian becomes "diagonal":
\begin{equation}
    H_{Harmonic}=\sum_{\lambda} \frac{P_{\lambda}^2 }{2} 	+  V_0 + \frac{1}{2}\sum_{\lambda}\omega_{\lambda}^2 U_{\lambda}^2 
\end{equation}{}
The transformation is the one that has diagonalized the rescaled, symmetrized force constant matrix: $\tilde{\Phi}_{i\alpha,j\beta}=\Phi_{i\alpha,j\beta} /\sqrt{m_i m_j}$. In other words: 
\begin{equation}
    \sum_{j,\beta} \tilde{\Phi}_{i\alpha,j\beta}\,  \epsilon_{j\beta,\lambda}= \omega_{\lambda}^2 \, \epsilon_{i\alpha,\lambda}
    \label{eigenmodes}
\end{equation}
with $\sqrt{m_i} U_{i\alpha}=\sum_{\lambda} \epsilon_{i\alpha,\lambda} U_{\lambda}$ and the orthogonality and completeness relations among the eigenvectors coming respectively from $\epsilon^{\dagger} \epsilon=\mathbb{I}$ and $\epsilon \epsilon^{\dagger}=\mathbb{I}$ \footnote{For a real symmetric matrix $\Phi$ the eigenvectors $\epsilon$ form an orthogonal matrix, while for a Hermitian one, they are unitary:
$\sum_{i\alpha} \epsilon_{i\alpha,\lambda}^* \epsilon_{i\alpha,\mu}=\delta_{\lambda,\mu}$;
$\sum_{\lambda} \epsilon_{i\alpha,\lambda}^* \epsilon_{j\beta,\lambda}=\delta_{i,j} \delta_{\alpha,\beta}$. }

This is the standard harmonic lattice dynamics theory, from which one can extract eigenvalues $\omega_{\lambda}^2$ (phonon frequencies squared) and eigenvectors (phonon polarization vectors). Normal coordinates play an important role in simplifying the mathematical formulation of thermodynamics of the harmonic model, since it reduces an interacting atomic Hamiltonian to N-independent or non-interacting harmonic oscillators, each having their own frequencies and polarization vectors. In other words, phonon modes can be thought of as independent harmonic oscillators. The translational invariance of the lattice structure leads to wavelike solutions and a dispersion relation between frequencies and wave-vectors (see appendix\ref{LD-EOM}). This harmonic model is accurate at low-enough temperatures and for heavy-enough atomic masses since the resulting atomic displacements remain small. If quantized, this model will also correctly predict the heat capacity at low temperatures. Anharmonic effects, however, have to be taken into consideration at higher temperatures. Also for cases like rare-gas/quantum crystals, where the zero-point motion is large, this approximation is not valid\cite{DeWette1967}. This can also be because for some interatomic potentials, the second-order term in the Taylor expansion of potential energy may not be positive definite. Conventional harmonic lattice dynamics does not work in this scenario as it leads to imaginary phonon frequencies. To make the structure stable, one needs to incorporate higher-order terms in the Taylor expansion of the potential energy.
The standard perturbation theory approach,\cite{Maradudin1962} where cubic and higher order terms in the Taylor expansion are treated as perturbation allows the calculation of renormalized phonon frequencies and their lifetime and temperature dependence.  The effectiveness of this approach depends on two major assumptions: 1- that the harmonic force constants form a positive definite matrix so that the resulting harmonic frequencies are real and positive numbers, and 2- that anharmonic effects are comparably smaller than the harmonic terms. In other words, the starting reference atomic positions are a local minimum of the potential energy and the dynamical displacements of nuclei should be strictly limited within the vicinity of equilibrium position for the perturbative method to work. However the small displacement assumption will not necessarily remain valid at high temperatures, especially at or near a phase transition nor for any other forms of instability in a lattice system. Even at low temperatures if the zero point motion is large enough, anharmonic terms can become important.

\begin{figure}[h]
	\centering
	\includegraphics[width=\linewidth]{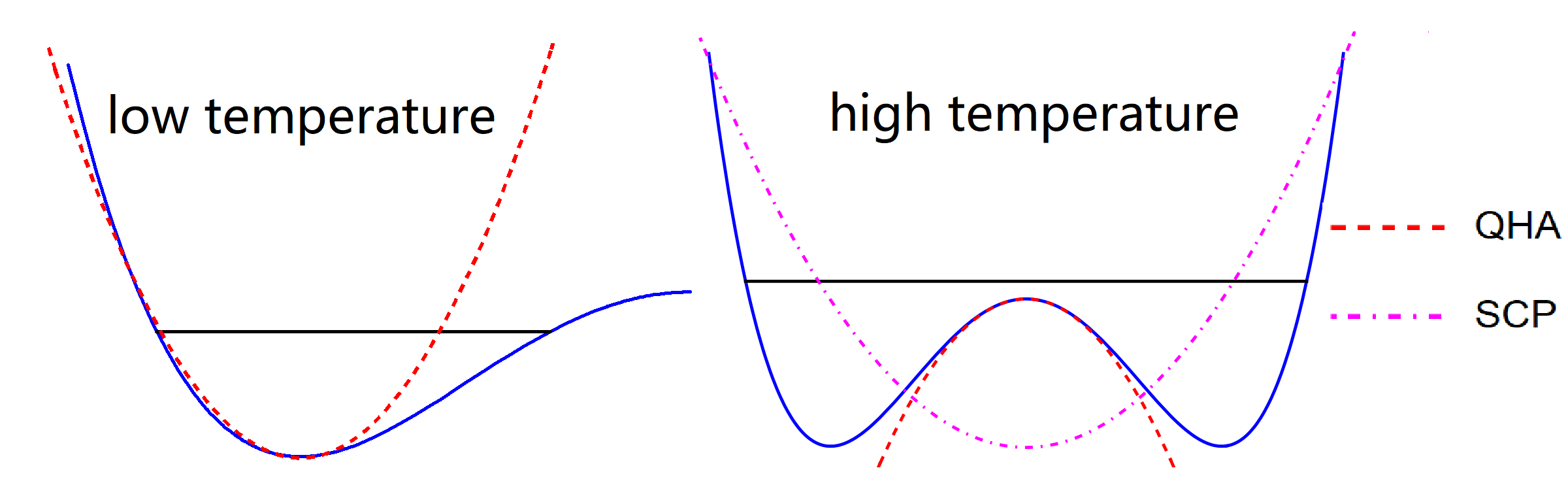}
	\caption{Intuitive illustration of Born's insight\cite{Hooton1955}. While the harmonic potential has a negative second derivative, the SCP one is positive definite and leads to "stable" phonons.}
	\label{fig:insight}
\end{figure}

The physics of the above problem can be explained as follows: for an atom in a crystal with low vibrational energy, within its oscillation scope around local minimum, the potential curve can be nicely fit with a positive curvature parabola. There are cases however, in which as the temperature increases, the atom may possess an energy level higher than the nearest potential hump, thus at that point the harmonic approximation will give a parabola pointing downward, which leads to imaginary phonon frequencies as shown in Fig.[\ref{fig:insight}], although on the average, the atom oscillates around the high-symmetry but unstable position. In order to overcome this difficulty, a new renormalized perturbation expansion later to be recognized as the self-consistent phonon theory has firstly been formally constructed by D.J.Hooton, 1955, in his journal article co-authored with his advisor, Max Born. \textit{Statistische Dynamik mehrfach periodischer Systeme}.\cite{Born1955}

\begin{figure}[h]
	\centering
	\includegraphics[width=\linewidth]{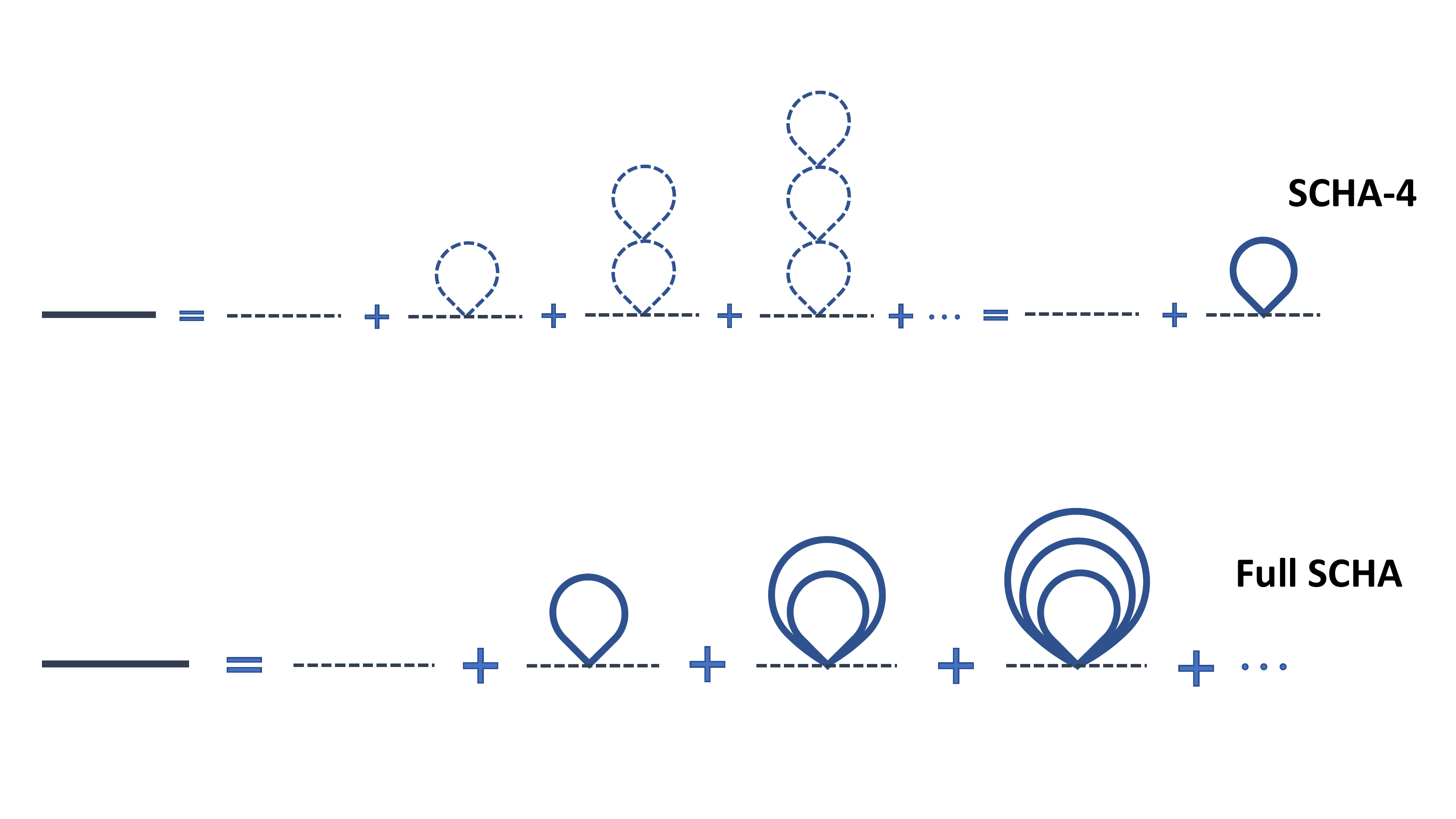}
	\caption{Feynman diagrams for the phonon propagator corresponding to the SCHA. Dashed lines represent the bare harmonic propagator while solid lines represent the self-consistent ones. Top is what we call SCHA-4 where only the quartic term is kept, and bottom corresponds to keeping all (even) derivatives in the expansion in eq. \ref{taylor}.}
	\label{scha-diagram}
\end{figure}

The basic concept of self-consistent harmonic approximation (SCHA also called by some authors SC1) is to fit the actual potential energy curve by an effective positive-definite parabola: to take  the thermally weighted average of the second derivative of the potential energy instead of taking the second derivative at the equilibrium position, which could be negative.
\begin{gather}
\Phi_{actual}\rightarrow \Phi_{SCHA}\equiv\, \langle \Phi_{actual}\rangle =\langle \Phi(R_{ij}+U_i-U_j)\rangle \\
=\Phi(R_{ij}) + \frac{1}{2} \langle UU\rangle \frac{\partial^2 \Phi }{\partial R^2}+\frac{3}{4!} \langle UU\rangle \langle UU\rangle \frac{\partial^4 \Phi } {\partial R^4}+...
\label{taylor}
\end{gather}
$\langle \Phi_{actual}\rangle $ means a thermal average over force constants with bond lengths fluctuating about their thermal equilibrium value. As a shorthand, $U$ symbolically represents $U_i-U_j$.
We will explicitly show the derivation of this equation as done originally by Boccara and Sarma\cite{Boccara1965}, in  appendix C.
By Taylor expanding the force constant in powers of atomic displacements $U_i^{\alpha}$ we see that we can calculate this average if all even-order derivatives are known, because displacements have a Gaussian distribution, and averages of the type $\langle U_i U_j...U_l \rangle $ can be exactly evaluated by using Wick's theorem\cite{Wick1950}. More formally, keeping in mind that the averaging is only over the dynamical parameters $U$, which has a gaussian distribution, and replacing for the sake of brevity $U_i-U_j$ by $U_{ij}$, this can be written as:
\begin{equation}
\langle \Phi(R_{ij}+U_{ij})\rangle=\langle e^{U_{ij}.\nabla_R} \rangle \Phi(R_{ij})= e^{\frac{1}{2} \langle U_{ij}U_{ij}\rangle: \nabla_R^2}  \Phi(R_{ij})  \nonumber
\end{equation}
The $\nabla_R$ operator, which is used in the exponent to express the translation, acts on the function $\Phi(R)$. This equation is formally equivalent to eq.\ref{taylor}.
The self-consistency comes in the evaluation of $\langle UU\rangle$ term which is expressed in terms of eigenvalues and eigenvectors of $\Phi_{SCHA}$ itself.
In terms of Feynman diagrams, this approximation can be described by the sum of the diagrams in Fig.\ref{scha-diagram}, where the four-, six- and eight-point vertices correspond respectively to the fourth, sixth and eighth derivatives of the potential energy with respect to atomic displacements and the dashed and solid lines correspond to the bare harmonic and self-consistent harmonic propagators (or Green's functions) respectively.    

 Hoerner has shown that this self-consistent perturbation approach is equivalent to the variational approach\cite{Horner1967} originally proposed by Boccara and Sarma\cite{Boccara1965}, which asks the question: what  are ``best" trial harmonic force constants that minimize the system's variational free energy? In a mathematical sense, we are using a quadratic ``trial" potential $V_{trial}$ to best fit the ``actual" potential according to the actual displacements taking place at a given temperature. The equivalence is also shown by Gillis in 1968 \cite{Gillis1968}, who uses the variational formulation, and shows the best trial force constants can in fact be expressed in terms of sum of even-order force constants as shown in Eq.\ref{taylor}. 
We will provide a derivation of these equations using the variational approach later in the text. 

\subsection{Implementation example of the variational approach}\label{example}
\subsubsection{Variational formulation}
%The full SCI theory\cite{Gillis1968} may be derived from diagrammatic Green's functions theory, as well as from a variational formulation. 
We will first illustrate the variational formulation with a simple example where the free energy of a one-dimensional anharmonic oscillator is minimized in accordance with the Bogoliubov inequality (BI). BI provides a strong tool that provides the mean-field equations treating a system at finite temperature. The Bogoliubov inequality upon which the variational formulation is based, is as follows:
\begin{equation}
F_{actual} \leqslant F_{0}+\langle H_{actual}-H_{trial}\rangle_0
\label{eq:BI}
\end{equation}
Here $F$ is the actual (exact) free energy and $H$ is the actual Hamiltonian, $H_{trial}$ is any trial solvable Hamiltonian, meaning for which the density matrix is known and the free energy $F_0$ is readily calculated. The sharp brackets $\langle \; \rangle_0$ mean thermal average taken with a density matrix defined with $H_{trial}$:
\begin{gather}
    \langle A \rangle_0= {\rm Tr} \,\rho_{trial} A \\ 
    \rho_{trial}=\frac{e^{-H_{trial}/k_BT} }{Z_0} \\
    Z_0={\rm Tr}\, e^{-H_{trial}/k_BT}; F_0=-k_BT \,{\rm ln} \,Z_0
    \label{trial} 
\end{gather}
    
Thus the effect of temperature is introduced through the averaging procedure with respect to a known density matrix or distribution function. 

The best choice of trial Hamiltonian is the one that leads to the lowest value of the right hand side of eq. \ref{eq:BI}, which is therefore the closest possible to the true free energy. So in practice one chooses a solvable trial Hamiltonian with some parameters which will be determined by minimizing the right-hand side of eq. \ref{eq:BI}. This is the meaning of the variational approach.

In SCP and Feynman's example introduced below, the trial Hamiltonian $H_0=H_{trial}$ always takes the form of a harmonic oscillator which is exactly solvable. After applying the variational formulation, the trial mass can be shown to be the same as the real mass of the particle. The $\langle H_{actual}-H_{trial}\rangle_0$ can therefore be replaced by the potential energy part $\langle V_{actual}-V_{trial}\rangle_0$. In his book\cite{Feynman1972}, Feynman gave an elegant proof of Eq.[\ref{eq:BI}] by using the Baker-Campbell-Hausdorff expansion relating the product of two exponentials to the exponential of the sum and correction terms involving commutators. It can also be derived from simple matrix and algebra and perturbation theory by using the Cauchy-Schwarz inequality\cite{Prato1995}. 

\subsubsection{The single-oscillator case}
Consider a single particle in a general bounded potential $V_{actual}(U)$ with $U$ being the dynamical displacement from the equilibrium position. We adopt a trial Hamiltonian with a displaced equilibrium position $u_0$ (in order to incorporate thermal expansion) and a displaced oscillation frequency $\omega_0$, both of which need to be calculated by minimizing the right-hand side of eq.\ref{eq:BI} with respect to these two parameters. Feynman showed that the minimization results in two concise equations\cite{Feynman1972}: 
\begin{eqnarray}
	\partial F_{trial}/\partial u_0=0 &=>& \langle V_{actual}'(U)\rangle_0 = 0  \label{eq:1} \\ 
	\partial F_{trial}/\partial \omega_0=0 &=>& \langle U\,V_{actual}'(U) \rangle_0 = \langle P^2/m\rangle_0 \label{virial}
\end{eqnarray}
Eq.\ref{eq:1} is the statement that at equilibrium, the net average force on the particle should be zero. It is derived from the minimization with respect to the displaced equilibrium position $u_0$ . Eq.\ref{virial} is more interesting since it is a statement of the virial theorem. It is derived from the differential of the trial free energy with respect to the effective oscillator frequency. But more generally, the virial theorem can be derived from a consideration of the change in the total energy under volume or more generally under a length scale change. Expressing that this change, which is essentially the pressure, is zero at equilibrium, one reaches Eq.\ref{virial}.

Below, we will use a toy model to showcase how those formulas work. For the sake of simplicity and to reduce heavy notations, we choose a %one dimensional monoatomic oscillator chain as shown in  Fig.[\ref{fig:monoatomic}] with the anharmonic cubic and quartic terms in the potential energy $V_{actual}$. 
particle of unit mass in an anharmonic potential with cubic and quartic terms in the potential energy $V_{actual}$.
%\begin{figure}[h]
%	\centering
%	\includegraphics[width=\linewidth]{image2}
%	\caption{Monoatomic 1D Chain\cite{Cottam2015}}
%	\label{fig:monoatomic}
%\end{figure}
\begin{align}
	V_{actual}=&\frac{1}{2}\omega^2 \, U^2+\frac{1}{6}a\, U^3+\frac{1}{24}b\, U^4 \\
	V_{trial}=&\frac{1}{2}\omega_0^2 \, (U-u_0)^2
\end{align}
$V_{actual}$ is the toy model potential with $U(t)$ being the atomic dynamical displacement from its equilibrium position ($U=0$). Both a and b, which are known parameters, can be thought of as anharmonic spring constants. $V_{trial}$ is the trial potential with variational parameters $\omega_0$ and $u_0$ which we want to determine from free-energy minimization. The former can be seen as the 'renormalized' phonon frequency and the latter is related to the thermal expansion. Since the trial system is purely harmonic, one can find the exact expression for $F_{trial}$ and $\langle V_{trial} \rangle$ as shown in Appendix Eq.(\ref{eq:harmonicF}) and Eq.(\ref{equipart}) respectively. We can substitute in and rewrite the Bogoliubov inequality as:
\begin{equation}
	F_{actual}\leqslant \frac{1}{\beta} {\rm ln[2 sinh}(\frac{\beta \hbar\omega_0}{2})] + \langle V_{actual}\rangle - \frac{\hbar\omega_0}{4}{\rm coth}(\frac{\beta\hbar\omega_0}{2})
\end{equation}
where $\beta = 1/k_BT$ and $k_B$ is the Boltzmann constant. The right side of this equation, the variational free energy, represents the upper bound of actual free energy. One can directly take its first derivatives with respect to the two variables $\omega_0$ and $u_0$ to set up the following two coupled equations:
\begin{gather}
	\frac{\partial \langle V_{actual}\rangle_0}{\partial u_0} = 0 \\
	\frac{1}{\hbar}\frac{\partial \langle V_{actual} \rangle_0}{\partial \omega_0}+\frac{\beta\hbar\omega_0}{8} \frac{1}{{\rm sinh}^2(\beta\hbar\omega_0/2)}=0
\end{gather}
Note that the thermal average is with respect to the distribution function of $H_{trial}$. the potential energy term $\langle V_{actual} \rangle_0$ involves powers of $U$ which can readily be calculated.
To simplify the calculation of the above derivatives, one can use the following results which hold for the variable $U-u_0$ having a Gaussian distribution: 
\begin{eqnarray}
\langle (U-u_0)^{2n+1} \rangle_0 &=& 0 \\ \nonumber
\langle (U-u_0)^{2n+2} \rangle_0 &=& (2n+1) \langle (U-u_0)^{2n}\rangle_0 \langle (U-u_0)^2 \rangle_0  \\ \nonumber
\langle (U-u_0)^2 \rangle_0 &=& \frac{\hbar }{ 2 \omega_0} {\rm Coth} (\frac{\beta \hbar \omega_0}{2}) 
\end{eqnarray}
The two variational parameters $(\omega_0,u_0)$ thus can be found by either numerically solving the above two coupled equations, or analytically with some approximations and in some specific limits. 
%or in a self-consistent numerical scheme by iterative methods such as conjugate gradients. 
If we were to use the small oscillation limit $u_0 \ll 1$ and high temperature limit $\omega_0 \ll k_BT/\hbar$, the transcendental equations can be simplified and be solved analytically:
\begin{gather}
	u_0 \approx -\frac{a T}{2\omega^4}+O(\frac{1}{T}) \nonumber \\
	\omega_0^2 \approx \omega^2 - \frac{bT}{2a\omega^2} + O(\omega^{-4})
\end{gather} 
The classical thermal expansion for 1D harmonic chain\cite{Ernest1956} is thus recovered. With more accurate calculation of $u_0$ and $\omega_0$, the Helmholtz free energy can thus be computed and plotted as a function of temperature. For more complicated systems such as ferroelectrics, one could investigate possible structural phase transitions by looking at the contour plots of the free energy in the $(T,u_0)$ plane. Above is just the simplest implementation of the self-consistent harmonic approximation SCHA (or SC1).  In a real lattice system, various computational techniques based on density functional theory(DFT) or molecular dynamics are practiced to approximate high-order FCs. The 'Gaussian smeared' $\langle V_{actual}\rangle_0$ is a notable feature for self-consistent theories since the trial Hamiltonian is quadratic and displacements about the equilibrium sites have a Gaussian distribution.

\section{Overview: Historical Development}\label{history}

Now that the main ideas behind the SCP have been clarified with an example, we proceed to give an overview of the method building on the harmonic theory.
%historical developments of this theory.

%(notes)
%-from 1958, Born and Hooton model\\
%-1966~1970, Horton, Choquard, Koehler, Werthammer, Gillis, Glyde; mostly applied to Leonard-Jones model (rare gas atoms) where $\phi(r)$ is known exactly\\

Attempts such as \textit{ab initio} molecular dynamics (AIMD)\cite{Car1985}, path-integral molecular dynamics (PIMD)\cite{Ceperley1995} or imaginary-time path-integral Monte Carlo (PIMC) to treat quantum anharmonic systems are either computationally intensive (PIMC, PIMD), or not properly including quantum effects (AIMD) or limited to equilibrium properties calculation, which severely limits their scope of applicability. 
Thus many approaches in order to overcome these limitations have been developed by different groups based on the renormalized perturbation expansion method, later to be known as the self-consistent phonon (SCP) theory or self-consistent harmonic approximation(SCHA) firstly introduced by Born\cite{Born1955} and then his student, Hooton\cite{Hooton1955,Hooton1958}, who in his second paper introduced a variational formulation of the problem. 
In the context of phonon softening and phase transitions, the variational approach was also adopted by Boccara and Sarma in 1965\cite{Boccara1965}, where they defined a harmonic Hamiltonian with displaced coordinates and effective force constants, to be determined by minimizing the free energy. 
Later, in 1966, Koehler introduced a similar approach \cite{Koehler1966} applied the self-consistent harmonic phonon approach to solid Ne at zero temperature, where, due to large zero-point motions, atoms were probing anharmonic regions of the potential energy. His results produced a lower value of the ground state energy\cite{Koehler1966} compared to those of Bernades\cite{Bernardes1958}, and Nosanow and Shaw\cite{Nosanow1962} who used an uncorrelated Einstein model to describe localized excitations in solid Ne. 
In that letter, his method was basically identical with quasi harmonic phonon approximation which he referred as 'self-consistent Hamiltonian', although the energy optimization is based on a set of uncorrelated Gaussian wave functions. He also claimed that additional feature of self-consistency was presented despite being inherently similar with Nosanow and Werthamer's paper\cite{Nosanow1965} in 1965 on sound velocity calculation.

SCP provides an alternative to the earlier perturbation theory approaches, in that it is a {\it mean-field theory} consisting essentially in replacing the harmonic force constants (or dynamical matrix) by their average over atomic displacements taking place at a given temperature. Naturally, this average is defined by anharmonic terms in the expansion of the forces in powers of atomic displacements. The "self-consistency" condition comes from expressing that the average over atomic displacements has to be performed with respect to the {\it renormalized} phonon modes and not the original harmonic ones. This theory can be derived in two different ways. One is from the diagrammatic perturbation method, where a series of diagrams are added to infinite order, and the other is from a variational  
approach wherein the free-energy of the crystal is minimized with respect to some {\it effective} force constants, and other variational parameters if needed, in order to obtain the self-consistent or mean-field equations to be solved. 

In 1965, Ranninger\cite{Ranninger1965} published a work based on diagrammatic perturbation theory, which he attributed to Choquard. Choquard's book\cite{Choquard1967} on the subject appeared two years later in 1967. He had generalized the method to second-order, which was a more complete theory since it included phonon damping.  
Later, by implementing a selective re-summation of diagrammatic perturbation theory, Horner showed that the two formulations were identical \cite{Horner1967}. The free-energy minimization is however more elegant and on solid grounds, and more amenable to future extensions (e.g. to describe phase transitions) if need be.

The first complete SCP theory was however first worked out as a mean-field theory by Gillis et al. in 1968\cite{Gillis1968}. It offered a starting point for quantum crystals studies. Techniques such as Green's function methods, variational approach or diagram summation all lead to the full SCHA  scheme. Gillis and Werthammer's SCHA (or SC1) theory are established based on several authors' previous work as well as Koehler's computational implementations.

Gillis, Werthamer and Koehler's SCHA theory (see Fig.\ref{scha-diagram}) was limited due to the omission of odd derivatives of the potential. An improved self-consistent phonon approximation (ISC) was introduced by Goldman, Horton and Klein in 1968\cite{Goldman1968}. The cubic term has been directly added as a second-order correction term to the Helmholtz free energy. This is done after the free-energy minimization, which has produced effective phonon frequencies. This was shown to correspond to the leading correction term in Choquard's first-order self-consistent equations.
With this addition, the free energy is not variational anymore but may well exceed the first-order theory in accuracy. This method had been carefully tested on a selection of rare gases, with the adoption of a phenomenological nearest-neighbor central-force potential. The results show that quantities such as the high temperature heat capacity $\mathcal{C}_p$ agrees best with the experiment, compared to the SCHA and simple perturbation theory of Feldman and Horton\cite{Feldman1967}.  They also agree well with this perturbation theory at roughly 1/3 of the melting temperature, thus the validity of this method holds at low temperature for rare gases.

One can not avoid Werthamer's contribution while discussing early SCP development. After Horner's paper in Februray of 1967, Werthamer published a paper on phonon frequencies and thermodynamic properties of crystalline bcc He calculations, co-authored with Wette and Nosanow, in May 1967\cite{DeWette1967}. Regardless of a more detailed mathematical formalism, their method was similar to Koehler's: using the time-dependent Hartree approximation together with the results of variational calculations of the ground-state energy using correlated trial wave functions. The advantage of the Hartree approximation is that it yields a self-consistent solution in which the particles are indeed localized. This is a convenient way to introduce the structure of the crystal into the calculation. Both short-(atomic repulsion) and long-range(phonons) correlations are considered. They assumed that with external imposed disturbances, the system wave function can still be factorized for all times into a product of single-particle functions. The phonon frequencies are defined as poles of the response function, just like the collective excitations in the electron gas. For short-range correlations, they adopt the results of variational calculation of the ground-state energy using cluster-expansion techniques. Half a year later, Werthamer along with Gillis and Koehler\cite{Gillis1968} constructed the complete self-consistent formalism, later to be known as SC1 theory. Nevertheless, their formulas for the self-consistent phonons are equivalent with those of Ranninger, Choquard and Horner. In Werthamer's 1970 paper\cite{Werthamer1970}, he organized the previous authors' work in a single functional variational formalism and extended it to not only give the leading cumulants of the free energy, but also the first-order thermodynamic derivatives. Thus a phonon dynamical model with adjustable parameters to fit the lower part of excitation spectrum of the model system as closely as possible, was constructed. Werthamer also took damping of phonons into account, by considering a trial action with force constants non-local in time. 

Once actual free energy is specified by this scp approximation, an equation then can be given to calculate the pressure for a particular choice of lattice constants; evaluations of isothermal elastic constants, specific heats and thermal strain were also given in this paper. Werthamer then moved toward the second-order term, which gives the leading contribution to the phonon damping rate and also produces a frequency shift connected with the damping through the Kramers-Kronig relations. %These formulas that were given by Werthamer are complicated but also contain some significant features, such as incorporation of decay process. The traditional anharmonic perturbation theory prescribes a phonon decay only into two other phonons as a result can be verified in his scheme. Multiple phonons processes contribute notably to the anharmonic terms especially at or above the Debye temperature. 
With the inclusion of second-order term, now odd numbers of derivatives of the potential enter into SCP theory. %Also from straightforward algebra that the zero-frequency long-wavelength limit of the phonon dynamical matrix can lead to a generalization of the isothermal elastic constants. 
A substantially simplified version for second-order approximation approach had been followed by Goldmen\cite{Goldman1968} and Koehler\cite{Koehler1969} in their numerical calculations, with a slightly improved agreement with experiment over the first-order phonon spectrum.

Later practices on full second-order theory led to divergence difficulties, with the physical reason explained by Horton in his book with Maradudin: \textit{Dynamical properties of solids}\cite{Horton1974}. He showed that the Gaussian pair distribution function that occurs in the thermal averaging penetrates too far into the repulsive core of the potential. The fully established SCP theory offers a practical approach for computing structural and dynamical properties of a general quantum or classical many-body system that also incorporates anharmonic effects and has advantages over standard harmonic approximation, although convincing demonstration of its accuracy had been limited to rare gases in its early history.

In 1973, Samathiyakanit and Glyde offered an alternative path integral formulation of the anharmonic lattice problem\cite{Samathiyakanit1973}. Their derivation uses a trial harmonic action. The exact partition function is then expanded in cumulants about the trial harmonic partition function. Successive orders of the self-consistent theory are then obtained by keeping successive cumulants and requiring that their contribution to the crystal dynamics vanishes. The method provides a description of both the vibrational dynamics and thermodynamics. This procedure is similar to Choquard's who also used a cumulant expansion method. Their theory also agrees with Werthamer's\cite{Werthamer1970} up to first order, but not to second order where it was claimed the latter theory has omitted some extra terms. 

\section{Modern interpretations and implementations} \label{modern-implementations}
Theory development for anharmonic lattices reached its limit in the 70's and it was implemented for rare-gas atoms where the interparticle potential is of two-body type and relatively well-known. With the advent of density functional theory methods, which enabled accurate calculation of many-body forces and force constants in real materials, these theories have been revived in the past decade. The major difference among different developments and implementations have been in two directions: A-the selection and extraction of force constants, and B- the sampling of phase space in order to calculate the thermal averages.

\subsection{Selection and extraction of force constants}
There are essentially three different approaches to obtain the force constants: a) the reciprocal space method, which is based on the density functional perturbation theory (DFPT) and the $2n+1$ theorem\cite{baroni1987green,Gonze1989,Debernardi1995}; and b) the real-space method which is based on finite displacements of atoms in a supercell\cite{Vanderbilt1984,Narasimhan1991,Parlinski1997,Esfarjani2008,ShengBTE2014,Togo2015,Tadano2018,Hellman2013-1}. 

\subsubsection{Force constant from DFPT (reciprocal-space formulation) }

A predictive theoretical scheme that originates from first-principles approach, with its emphasis on anharmonic decay of phonons was developed by Debernadi et al. in 1995.\cite{Debernardi1995}. This method is based on density-functional perturbation theory (DFPT)\cite{Gonze1989,Giannozzi1991} that worked well for semiconductors, and the use of $2n+1$ theorem, which, in particular, states that cubic force constants, which are third derivatives of the potential energy, can be expressed in terms of only first derivatives of the electronic wavefunctions. The method was then implemented to the calculation of phonon lifetimes in diamond, Si and Ge. The importance of this method was that it offered a reliable third-order force constant calculation scheme that could demonstrate anharmonic decay of phonons in semi-conductors. A comparison between computational and experimental data was provided to support the methodology. The advantage by using density-functional perturbation theory over standard lattice-dynamical calculations in harmonic approximation is that the former one offers an accessible way in terms of computational cost, to describe anharmonic phonon lifetimes, especially considering it may not be easily evaluated in experiments.

A decade later,  in 2007\cite{Broido2007}, Broido came up with a novel method, also employing DFPT, to compute intrinsic lattice thermal conductivity. Due to anharmonic phonon-phonon scattering and difficulties in accurately describing interatomic forces, a theoretical approach to predict lattice thermal conductivity  has been difficult since Peierls firstly introduced the phonon Boltzmann equation. Broido's treatment involved no adjustable parameters with the only input required being anharmonic FCs, which are determined from first-principles using DFPT. In the three-phonon process, the scattering matrix element is as following:
\begin{widetext}
    \begin{equation}
	V_{q_1\lambda_1,q_2\lambda_2,q_3\lambda_3} =\frac{1}{N} \sum_{1,2,3} \Psi_{R_1 \tau_1,R_2 \tau_2,R_3 \tau_3} \,
	e^{i (q_1\cdot R_1+q_2\cdot R_2+q_3\cdot R_3)}  	 \times\frac{\epsilon_{R_1 \tau_1,q_1\lambda_1} \,\epsilon_{R_2 \tau_2,q_2\lambda_2} \,\epsilon_{R_3 \tau_3,q_3\lambda_3}}
	{\sqrt{m_{\tau_1}m_{\tau_2}m_{\tau_3}}}  	 
    \end{equation} 
\end{widetext}

where $R$ labels a unit cell and $\tau$ labels an atom in the unit cell, of mass $M_{\tau}$, while the $\Psi_{1,2,3}$ is the third order anharmonic interatomic force constant for the indicated triplets of atoms, and $W$'s are polarization (or eigen-) vectors of the dynamical matrix as defined in eq. \ref{eigenmodes}. They form a unitary transformation from the atomic and cartesian $R \tau$ basis to the normal mode $q \lambda$ basis, and can be written as the product $e^{iq.R} \epsilon_{\tau, \lambda} (q)$ where the latter are the so-called polarization vectors (on atom $\tau$) of mode $\lambda$ at wavevector $q$. This scattering matrix element term is used to determine three-phonon scattering rate using Fermi's golden rule. From the Peierls-Boltzmann equation (PBE), nonequilibrium distribution functions can be solved via an iterative approach, and the lattice thermal conductivity tensor can be obtained. In this theoretical calculation scheme, the only input required are force constants: harmonic $\Phi_{1,2}$ and anharmonic $\Psi_{1,2,3}$. In that sense, it is crucial to have accurate FCs. In the DFPT approach their Fourier transform is directly calculated in the reciprocal space with the help of the $2n+1$ theorem. The $2n+1$ theorem\cite{Gonze1989} states that the $2n+1^{th}$ derivatives of the total energy can be obtained from the knowledge of the $n^{th}$ derivatives of the wave functions. Broido worked with a method firstly developed by Deinzer in 2003\cite{Deinzer2003} which calculates Fourier coefficients $V$ of the cubic force constants $\Psi$ at arbitrary wavenumbers from a Fourier interpolation. The method was first tested on silicon and germanium, and provided an outstanding agreement with experimental measurements in terms of lattice thermal conductivities.  However this framework is not able to compute fourth order terms, which would require the second derivatives of the wavefunctions. The alternative to higher-order DFPT, is to use a finite difference on low-order DFPT force constants. This was first implemented by Rousseau and Bergara in 2010\cite{Rousseau2010} who used the frozen-phonon approach to extract the cubic FCs from finite difference of dynamical matrices. Their purpose was to see the effect of cubic anharmonicity on the strength of the electron-phonon coupling in aluminum hydride superconductors.
The DFPT approach could be applied to various materials within a considerable scope of temperature, without introducing any extra parameters, if thermal expansion is negligible. However, since the FCs are calculated in reciprocal space, this approach becomes impractical for bulk materials with very large primitive cells as we will clarify later.

\subsubsection{Extraction of  force constants in real-space}

{\bf * The frozen phonon method }
The first calculations of phonon spectra from DFT originated in the 1980's, and the method to extract harmonic force constants was the so-called frozen-phonon method. This is a finite difference method in which atoms in a supercell, which we label here with $R\tau$, are moved according to a frozen phonon mode of wavenumber $k$ commensurate with the superlattice, and forces on atoms are calculated. In essence the displacement $\bm U_{R\tau}(k) =\bm \epsilon_k \rm cos k.(R+\tau) $ is imposed and the force $\bm F_{R'\tau'}(k)$ is calculated on all atoms in the supercell. The ratio between the two gives directly the force dynamical matrix elements at that wavenumber $k$ as we must have $\bm F_{R'\tau'}(k)=-\sum_{R\tau} \Phi_{R'\tau',R\tau}(k) \bm U_{R\tau}(k)$. Since there is only one term in this sum, the force constant $\Phi$ is obtained by taking the ratio of the force to displacement. This formula does not have finite difference if the atoms are originally relaxed so that the force for $\bm U_{\tau}=0$ is zero. This must however be done for many commensurate $k$ vectors and a Fourier transformation will give the force constants in the real space from which the dynamical matrix at any arbitrary $k$ vector maybe computed. 

{\bf * Small displacement method } 
This method is implemented in the codes PHONON\cite{Parlinski1997}, PHON\cite{Alfe2009}, ShengBTE\cite{ShengBTE2012,ShengBTE2014} and PHONOPY\cite{Togo2015}. 
In 1997, Parlinski\cite{Parlinski1997} presented a direct approach for phonon dispersion calculation, using \textit{ab initio} force constant method. The method uses Hellmann-Feynman theorem to calculate forces on displaced atoms in a supercell, and extracts all the harmonic force constants definable in that supercell. The displacements are not along a phonon mode polarization, and can be reduced if symmetry of the crystal is used. Effectively, this method assumes the cutoff range of interactions equals the supercell size. As a result, it reproduces the dynamical matrix at $q$ points commensurate with the supercell. For instance, if the supercell is of size 2x2x2, the dynamical matrix is exact at the zone center and zone boundaries which correspond to $G_1/2, G_2/2$ and $G_3/2$. Then the dynamical matrix and phonon dispersion are obtained at an arbitrary q-point by using the Fourier interpolation method. In this method, the dynamical matrices are calculated on a coarse q-mesh (2x2x2 in this example, corresponding to the second neighbor range in real space). They are then Fourier transformed and their Fourier coefficients are in effect the force constants, which will then be used to compute the dynamical matrix at any arbitrary q-point. The force constants are determined by considering symmetry-inequivalent displacements of atoms in the supercell.
This scheme was applied to calculate phonons in the cubic phase of $ZrO_2$ (zirconia), the Hellmann-Feynman force constants were acquired using \textit{CASTEP} program, adopting the  local-density-functional approximation(LDA) on a fcc supercell of 96 atoms. 
%Forces that were exact at several specific k points ($\Gamma,X,L,W...etc$) have been used for interpolation, 
The amplitude of atomic displacements had been limited to $u_0=\pm 0.010 a_0$ with $a_0$ being the lattice parameter %, as the only free parameter for optimization of total energy. 
In this method the range of force constants is determined by the supercell size.  Cubic terms are not considered in this approach either. In case of ionic materials, the Born charges $Z^*$ and the dielectric constant $\epsilon_{\infty}$ need to be calculated (from a separate DFPT calculation) and their contribution added as a non-analytical correction to the short-range part of the dynamical matrix. This correctly leads to the splitting of TO from LO modes\cite{Detraux1998,Parlinski1998b,Wang2010}.
 In such calculations, imaginary phonon frequencies indicate a lattice instability implying the structure is not at an energy minimum and there are displacements that can further lower the energy of the structure. 

In 2018, Parlinski's new paper\cite{Parlinski2018} formulated a method to model anharmonicity without actually computing the cubic force constants. In this method larger-amplitude symmetry-adapted atomic displacements in a super cell are used to compute various dispersions corresponding to various large amplitude displacements. The standard deviation in the obtained phonon dispersions will give an indication of the inverse lifetimes. 

{\bf * Inclusion of anharmonic FCs within a cutoff} This method is also implemented in ALAMODE\cite{Tadano2014c},  TDEP\cite{Hellman2011} and the lesser known, but equally performant \texttt{hiPhive} package developed in Erhart's group \cite{Erhart2019}. 

In 2008, Esfarjani and Stokes\cite{Esfarjani2008} were the first to introduce a method for extraction of higher-order force constants of in-principle any rank and up to any neighbor shell from first principles calculations in one or more supercells. This is in contrast to DFPT which can calculate up to cubic terms. If higher-rank terms were needed, major coding would be required to compute second derivatives of wavefunctions and the resulting fourth/fifth-order force constants. Another alternative within DFPT  would be to use a finite difference method to obtain the quartic terms from finite difference of cubic ones. 

In this method, the force on each atom in the supercell is expanded in powers of atomic displacements:
\begin{align}
\bm F_{R\tau}&=-\Pi_{\tau}-\sum_{R_1 \tau_1} \Phi_{R\tau,R_1\tau_1} \, \bm U_{R_1\tau_1} \nonumber\\
&-\sum_{R_1 \tau_1,R_2 \tau_2} \Psi_{R\tau,R_1\tau_1,R_2 \tau_2} \, \bm U_{R_1\tau_1}  \bm  U_{R_2 \tau_2} + ...
\label{fd}
\end{align} 
where, as before, $R$ refers to a primitive cell within the supercell and $\tau$ refers to an atom in the primitive cell, so that the pair $R\tau$ refers to an arbitrary atom within the supercell. The force constants $\Phi$ and $\Psi$ are the unknown parameters of the model. 

The methodology takes following steps: (1) symmetry operations of the crystal are calculated, and accordingly, the irreducible unknown force constants are identified. The remaining ones can be deduced from irreducible ones using symmetry operations $\Phi_{S(\tau),S(\tau')}^{\alpha\beta}= \sum_{\alpha'\beta'} S_{\alpha\alpha'} S_{\beta\beta'} \Phi_{\tau,\tau'}^{\alpha'\beta'}$ which simply says: to get the FC tensor of a ``rotated" bond, one simply needs to rotate the 3x3 tensor of the original bond;  (2) a cutoff distance is chosen for each rank of force constants, and thus the number of unknown force constants is fixed, (3) for appropriately chosen atomic displacements, the force on all atoms $\bm F_{R\tau}$ in the supercell is calculated using any first-principles method. Several sets of such displacements are chosen in order to generate enough ``equations" from which the unknown FCs can be deduced. (4) There are additional constraints of translational and rotational invariance which are imposed as additional linear equations the FCs $\Phi, \Psi, ...$ must exactly satisfy. These relations express the fact that if the crystal is translated or rotated by an {\it arbitrary} amount, the total energy should remain unchanged. Given that the force-displacement relation itself (eq. \ref{fd})  is also linear in the FCs, one ends up with an over-determined set of {\it linear} equations on FCs. These equations are solved using a singular-value decomposition (SVD) algorithm, which in essence means the difference between DFT forces and anharmonic model forces of eq. \ref{fd} above is minimized with respect to the unknown selected FCs. 
The difference between this method and what we called ``small displacements method" is that the harmonic FCs are also chosen to be non-zero within a cutoff neighbor shell, whereas the previously mentioned methods include all harmonic force constants that exist within the supercell, and in fact they can only recover the combination ${\tilde \Phi}_{0\tau,R'\tau'}=\sum_L \Phi_{0\tau+L,R'\tau'}$ where $L$ is a translation vector of the supercell. The latter method can however, in principle, extract all force constants independently if force-displacement data on several supercells of different size and shapes are provided. 
The major merits of this real-space method are: (1) The methodology can employ force-displacement data in one or many supercells of various sizes and shapes in order to handle longer-range FCs. (2) The method, as implemented, includes cubic and quartic anharmonic terms which plays a crucial role in phonon life-time calculation, crystal stability, and second-order SCP theory. This however can be extended to higher rank FCs as well. 

The choice of atomic displacements is important if not crucial in a correct and reliable determination of the force constants.  They can be obtained either by moving atoms of the primitive cell one by one along the cartesian directions, or from an MD simulation, or from random displacements of given magnitude, or if one wants to get effective force constants at a given temperature, one can sample the canonical ensemble at that temperature. The choice ultimately depends on the purpose of the simulations.
For a pure harmonic FC extraction in order to obtain the phonon spectra, one can generate a minimal set of small displacements, typically 1\% or 2\% of the bond length. Atoms can either be moved one at a time in each ``snapshot" or even collectively. Symmetry considerations can reduce the number of needed displacements. For instance in Si where both atoms are identical and the three cartesian directions are equivalent, it is enough to move only one Si atom along the $x$ direction, in order to get all the harmonic force constants. Collective displacements may also be made according to the irreducible representations of the space group of the crystal. 

Always convergence checks need to be performed with respect to the chosen cutoff range and also the supercell size. In the reciprocal space  (DFPT) approach, this is done by increasing the size of the q-mesh used for the DFPT calculations. This is reflective of the equivalent supercell size. For instance a q-mesh of 4x4x4 is equivalent to a supercell which is 4 times longer in each direction, and accordingly the range of the FCs goes to the 4th neighbor cell.

 The real-space method provided accurate prediction of the force constants of a purely analytical Lennard-Jones potential \cite{Esfarjani2008}, providing therefore a validation of the approach. This implies  first principles forces need to be well-converged in order to achieve an accurate fit.
 However there is no systematic approach for the choice of the supercell size and the cutoff range of the FCs. The recent work by Marianetti's group\cite{Marianetti2019} has however considered this important question and has a comprehensive discussion of this topic, and we refer the reader to this work for more details on how to adopt a systematic approach. 
 This scheme is automatically well-suited for lattice distortions, because for small distortions one can use the higher-rank force constants to predict the lower-rank force constant of the distorted structure using the Taylor expansion:
 $\Phi_{new}=\Phi+\Psi. \eta + ...$, where $\epsilon$ is the strain field and $\eta=\epsilon.R$ represents the corresponding atomic distortion.

%The real-space method has been implemented in the code ALAMODE developed by Tadano\cite{Tadano2018}. Similar versions have also been implemented in Sheng-BTE\cite{ShengBTE2014}. The code PHONOPY\cite{Togo2015} is an implementation of Parlinski's method outlined earlier, which uses Fourier interpolation of the dynamical matrix. The code TDEP developed by Hellman\cite{Hellman2011,Hellman2013-1,Hellman2013-2,Hellman2017,Hellman2018} which will be discussed later uses the same extraction principle. The \texttt{hiPhive} package developed in Erhart's group \cite{Erhart2019} is also based on the real-space method. 

At the end, we should note that recently a real-space formalism for linear response calculations (DFPT) is presented and applied to harmonic phonon calculations by Carbogno and Scheffler\cite{Scheffler2017}. 

\subsubsection{Discussion of difference regarding FC extraction techniques}

Before moving on to more recent self-consistent phonon calculation methods, we give a brief comparison between the real-space and reciprocal space methods.  The major two differences are: (1)computational load; (2)symmetry and invariance conservation

Some of the major contributors of \texttt{QUANTUM ESPRESSO}, Baroni et al., in their review article \textit{Phonons and related crystal properties from density-functional perturbation theory}\cite{Baroni2001} give a detailed comparison between the computational loads of DFPT and snapshot supercell DFT calculations. Suppose the range of FCs is $\mathcal{R}_{FC}$, thus the supercell should be of that size and the number of atoms $N_{at}^{sc}$ in such a supercell should be proportional to its cube: $N_{at}^{sc}\propto\mathcal{R}_{FC}^3$. In a single snapshot real-space supercell DFT calculation the computational cost is proportional to the cube of the number of plane waves $N_{pw}$. This number is decided by the supercell length $L$ and the minimum wavelength $\lambda_{min}$: $N_{pw}=(L^3/\lambda_{min}^3)$. Notice that the numerator $L^3$ as the volume of this supercell is proportional to the number of atoms in the supercell $N_{at}^{sc}$. Thus the single snapshot computation time is on the order of $\mathcal{R}_{FC}^9$. A complete FCs extraction using the real space method requires at least $3N_{at}$ snapshots, where $N_{at}$ is the number of atoms in the primitive cell, so the total computational load for real-space methods is on the order of  $N_{at}\times\mathcal{R}_{FC}^9$. On the other hand, the reciprocal space method based on DFPT requires the evaluation of the dynamical matrix on a q-mesh, where the interval between neighboring q points is $\Delta q \approx 2\pi / \mathcal{R}_{FC}$, so that the total number of $\bm{q}$ points $N_q$ is roughly $N_q \propto \mathcal{R}_{FC}^3$. As for the calculation of the dynamical matrix for each $\vec q$ $\mathcal{D}(\vec{q})$, the cost is on order of $N_{at}^4$. Such calculation for $\mathcal{D}(\vec{q})$ is performed over the irreducible first Brillouin zone, so a DFPT calculation will require a CPU time on the order of  $N_{at}^4\times\mathcal{R}_{FC}^3$. The  ratio of these two times is therefore:
\begin{equation}
	\frac{\rm CpuTime_{DFPT}}{\rm CpuTime_{SC}}=O(\frac{N_{at}^4\times\mathcal{R}_{FC}^3}{N_{at}\times\mathcal{R}_{FC}^9})=O(\frac{N_{at}^3}{N_q^2})
	\label{eq:load}
\end{equation}   
From Eq.[\ref{eq:load}], for systems with small unit-cells that requires a relative large q-mesh size, a DFPT implementation will work better. On the other hand, for systems with large primitive cell, the real space supercell approach will be more efficient. The size of the supercell or the range of FCs is for most materials on the same order of a nanometer or less, whether they have a large primitive cell or a small one does not matter. So in both methods the $\mathcal{R}_{FC}^6$ term is always the same.
Another point to keep in mind is that in DFPT method, which provides an exact equation for the FCs, translational and rotational invariances should be in principle satisfied. In practice, however,  due to numerical inaccuracies due to convergence and finite basis sets, these invariance relations can be violated. We have not seen any systematic study of this issue. In routine DFPT calculations, however, the ASR (or the acoustic sum rule coming from translational invariance) is enforced by other means, implying that such violations exist and are not necessarily small.  

\subsection{Sampling of the configuration space for effective theories at finite temperature}

The above approaches would in principle provide force constants at zero temperature or the ``bare" force constants. DFPT is a linear response theory, so it assumes atomic displacements are infinitesimally small, although atoms are not even moved in this calculation as this is a linear response. In real-space methods, displacements are finite but very small so that the contribution of higher rank terms becomes almost negligible compared to the harmonic ones and the Taylor series are convergent. In practical applications, however, one needs the value of these parameters at finite and even high temperatures. This can be achieved using several sampling methods. Different existing codes and approaches they use differ in their sampling approach.
At first sight, one may think that this problem is not so well-defined because when temperature and therefore atomic displacements become large, anharmonic terms start to contribute more, and it is not clear the Taylor series as we wrote in eq. \ref{taylor} would even remain convergent! One would really need a different expansion if displacements and the corresponding anharmonic terms are large. It turns out that the correct ``Taylor expansion" at high temperatures corresponds to the self-consistent phonon (SCP) theory! Assuming there is no atomic diffusion and atoms vibrate around their equilibrium site at all times, we need to use a mean-field theory to describe the vibration of atoms. Terms in the Taylor expansion need to be rearranged so that the series converge. If the harmonic term is replaced by an effective one, one then requires that the remaining part of the potential energy be zero on the average:
$ \langle V_{total}(U) \rangle = 1/2 \Phi_{eff} \langle UU \rangle $. Here the average is with respect to the equilibrium state defined by the {\it effective} normal modes; the corresponding density matrix is $\rho_{eff} = e^{-\beta \Phi_{eff} UU/2}/Z_{eff}$. We can readily see that the effective harmonic FC is a sum of only even order terms and averages of even powers of displacements. This was illustrated in 
 Fig. \ref{scha-diagram}. In this case, the remaining higher-order anharmonic terms have a smaller contribution since their mean is zero by construction. To illustrate this better, let us consider symbolically a Taylor expansion up to only 4th-order term and regroup some terms:
\begin{equation}
    V=V_0+\frac{1}{2} \sum \Phi U^2 + \frac{1}{4!} \sum \rchi U^4 
    \label{pe4}
\end{equation}
Requiring this to have an average equal to $1/2 \Phi_{eff} \langle U^2 \rangle$ implies
\begin{align}
\frac{1}{2} & \sum \Phi_{eff} \langle U^2 \rangle=\frac{1}{2}\sum \Phi \langle U^2 \rangle  + \frac{1}{4!} \sum \rchi \langle U^4 \rangle \nonumber \\
&= \frac{1}{2}\sum \Phi \langle U^2 \rangle  + \frac{1}{8} \sum \rchi \langle U^2 \rangle \langle U^2 \rangle  
\end{align}
Where we used properties of variable $U$ having a Gaussian distribution. This equation uniquely defines the effective harmonic force constants, which are obtained by taking the derivative of the above equation with respect to $\langle U^2 \rangle$:
\begin{equation}
    \Phi_{eff} = \Phi  + \frac{1}{2} \sum \rchi \langle U^2 \rangle
    \label{scha-4}
\end{equation}

We now see that even for large values of displacements $U$, the contribution of the second term involving only $\rchi$ in the regrouped version is  smaller because the potential energy has now the form:
\begin{equation}
    V=V_0+\frac{1}{2} \sum \Phi_{eff} U^2 + \frac{1}{4!} \sum \rchi U^2 (U^2-6  \langle U^2 \rangle )
\end{equation}
where the second term has zero average by construction.
So although the first version may be divergent, the second formulation seems to have better convergence properties. This is the essence of the SCP theory and the reason behind its success. So the FCs at finite temperatures acquire a new meaning which is different from their zero temperature version as derivatives of the potential energy evaluated at equilibrium position.  We are in essence changing the theoretical model, and need a different way to get its parameters.  

At a fixed temperature, the system is sampling the canonical ensemble, and atoms will have displacements of amplitude commensurate with that temperature. The methods we outline below are different ways of performing this sampling.  

%A comparison diagram can be referred below, among those techniques, in Esfarjani's and Mauri's method the symmetry and invariance relation have been explicitly imposed. For other methods whether those constraints have been met or not remains questionable. 
%\begin{widetext}
%	\begin{figure}[h]
%		\centering
%		\includegraphics[width=18cm]{image6}
%		\caption{Comparison diagram of IFCs extraction method}
%		\label{fig:IFCs}
%	\end{figure}
%\end{widetext}

\subsubsection{SCAILD} 

The first implementation of SCP using first-principles forces was reported by Souvatzis et al.\cite{Souvatzis2008,Souvatzis2009} and was labeled as SCAILD, which stands for  self-consistent ab initio lattice dynamics. This method was introduced to explain the vibrational entropy stabilization of group IV materials such as Ti and Zr at high temperatures in their BCC phase. In this method, atoms are displaced in a supercell and forces on them are computed in order to obtain an initial phonon dispersion, albeit with imaginary modes. From the eigenmodes at wavevectors commensurate with the supercell a set of displacements are derived. For these displacements, the forces on all atoms are then computed and Fourier transformed. New phonon frequencies are then derived from the relation between the DFT forces and polarization vectors:
\begin{align}
\omega_{k \lambda}&=\sqrt{-\bm F_k^{\tau} \, . \bm\epsilon_{k \lambda}^{\tau} / m A_{k \lambda}^{\tau} } \,\,\,\,\,\,\,{\rm since }\nonumber \\
\bm F_k^{\tau} &= - \sum_{\lambda} m \omega_{k \lambda}^2 \bm \epsilon_{k \lambda}^{\tau} A_{k \lambda}^{\tau} \,\,\,\,\,\,\,\,{\rm with }  \nonumber \\
A_{k \lambda}^{\tau}&=\pm \sqrt{\hbar (2n_{k \lambda}(T)+ 1)/2m_{\tau} \omega_{k \lambda}}
\end{align}
As can be seen, $A_{k \lambda}^{\tau}$ is the displacement amplitude of mode $k \lambda$ on atom $\tau$ at temperature T, which is included through the Bose-Einstein distribution function $n_{k \lambda}(T)$.
Everytime a new set of frequencies are obtained, they are properly symmetrized according to the symmetry of the kpoints:
\begin{align}
	\Omega_{k\lambda}^2&=\frac{1}{p_{k}} \sum_{S\in S(k)}\omega^2_{S^{-1} k\lambda} \nonumber\\
	\omega_{k\lambda}^2&=\frac{1}{N}\sum_{i=1}^{N}\Omega^2_{k\lambda}(i)
	\label{eq:restore_sym}
\end{align}
here $S(k)$ is the symmetry group of the wave vector $k$, $p_{k}$ is the number of elements of this group and $\Omega_{k\lambda}(i)$ are the symmetry restored frequencies at iteration $i$. The average of this distribution due to several iterations is taken as the definition of the phonon frequency according to eq. \ref{eq:restore_sym}. In this sum one may want to only include the last few iterations. 
Once a new set of frequencies are obtained in this way, atoms in the supercell are moved again with the updated amplitudes $A_{k \lambda}$ according to: 
\begin{equation}
    \bm U_{R\tau}=\frac{1}{\sqrt{N}} \sum_{k \lambda} A_{k \lambda}^{\tau} \,\bm\epsilon_{k \lambda}^{\tau}\, e^{i k.R}
\end{equation}
This process is repeated until frequencies converge. Also notable is the fact that the eigenvectors are not updated during the iterations.
This method was later applied by Zhang et al. to treat the stability of bcc and fcc phases of Tungsten at high temperatures\cite{Zhang2017}.

\subsubsection{TDEP} 
 The temperature-dependent effective potential method, aka (TDEP) was proposed by Hellman et al. in 2011 \cite{Hellman2011}. This method used the atomic configurations from a molecular dynamics run and fitted the DFT forces to an effective harmonic model. 
 They used the symmetry properties of the crystal to reduce the number of FCs and performed a least square fit to extract the inequivalent FCs, in a similar spirit to the work of Esfarjani and Stokes \cite{Esfarjani2008}. 
 Rotational invariance constraints did not seem to have been imposed. This might be fine for a purely harmonic model, but has consequences if anharmonic FCs also need to be included, as rotational invariances relate the two sets. 
 The main difference with the latter method is that the effective FCs are temperature dependent as the MD snapshots were from a constant temperature run, and as such, the TDEP method, similar to SCAILD gives the best harmonic FCs, as effective couplings at a given temperature and can provide effective phonon dispersions at that temperature. The method was validated on BCC phases of Li and Zr crystals. 
 Two years later, in 2013, Hellman et al. extended their method to include free energy calculations\cite{Hellman2013-1}  by including anharmonic corrections to the free energy through thermodynamic integration. A later work that year included temperature-dependent cubic anharmonic FC extraction\cite{Hellman2013-2} which was applied to Si and FeSi. 
 In this work the effective harmonic and cubic force constants were extracted simultaneously from the minimization of the difference between the DFT forces and the anharmonic model forces, similar to the original work. 
 Due to the high computational cost on initialization of canonical ensemble at temperature T, and length of an ab initio molecular dynamics (AIMD) simulation, in a later work in 2018, Hellman et al. employed a formulation of atomic displacements sampling the canonical ensemble, proposed originally by Estreicher\cite{Estreicher2006} and called it ``stochastically initialized temperature-dependent effective potential" (s-TDEP) \cite{Hellman2018}. The following equation was used to generate random displacements modeling the canonical ensemble at temperature T. This requires the knowledge of the phonon frequencies $\omega_{k\lambda}$ and eigenmodes $\bm{\epsilon}_{k\lambda}^{\tau}$, which are initially deduced from minimal displacements in the same supercell:
\begin{align}
	\bm{U}_{\tau}&= \sum_{k\lambda} \sqrt{\frac{\hbar (2n_{k\lambda}(T) +1) }{2N m_{\tau} \,\omega_{k\lambda} }}  \sqrt{-2{\rm ln}\zeta_1 } \,{\rm sin}(2 \pi \zeta_2)\, \bm{\epsilon}_{k\lambda}^{\tau} \nonumber \\
%	\bm{\dot{U}}_{\tau}&=\sqrt{\frac{2k_B T}{m_{\tau}}}\sum_{k,\lambda}\sqrt{-{\rm ln}(1-\zeta_{k\lambda})}{\rm cos}(\omega_{k\lambda}t+\phi_{k\lambda})\bm{\epsilon}_{k\lambda}^{\tau} %e^{ik.R}
	\label{eq:canonical}
\end{align}
where $N$ is the number of k-points, $\tau$ refers to an atom {\it in the supercell}, unlike in previous notations, and $\zeta_1,\zeta_2$ are random numbers chosen uniformly in $[0,1]$ so that the product $\sqrt{-2{\rm ln}\zeta_1 } \,{\rm sin}(2 \pi \zeta_2)$ has a normal distribution of mean 0 and standard deviation 1. 
%phase $\phi_{k\,\lambda}\in[0,2\pi)$ defines a random value for kinetic and potential energy, $\zeta_{\vec{q}\,\lambda}$ is a uniformly distributed random number between 0 and 1 producing the Box-Muller transform. 
This feature saves major CPU time as AIMD simulations to reach thermal equilibrium and several uncorrelated snapshots was the most time consuming part of the work. Another disadvantage of the MD method is that it treats ions classically, and therefore the atomic snapshots generated from MD are a valid description of the canonical ensemble only at temperatures above the Debye, whereas eq. \ref{eq:canonical} includes the effect of zero point vibrations at low temperatures.
Assuming the snapshots used to extract FCs are correctly generated, this scheme does not require self-consistency as in the SCP theory, and includes implicitly through the DFT force, all other interactions with phonons, namely that of electrons (assuming the adiabatic approximation). 

\subsubsection{SCHA-4}
The SCP theory to lowest order will only contain the quartic term in the potential energy, similar to the example we used in eqs. \ref{pe4}. For this reason, we will refer to it as SCHA-4. In terms of Feynman diagrams, it only includes the first two diagrams in Fig. \ref{scha-diagram}. In this theory, the effective harmonic force constants, as we showed in eq. \ref{scha-4}, can be shown to be: $\Phi_{eff} =\Phi+ \rchi \langle UU \rangle /2$, where the average is with respect to the self-consistent ground state defined by $\Phi_{eff}$. This approach combined with first-principles was first applied by Vanderbilt et al.\cite{Vanderbilt1984,Vanderbilt1986} to the case of carbon, silicon and germanium, where FCs were obtained using the frozen phonon approximation.
More recently, Rousseau et al. applied this theory in 2010  to study the effect of cubic anharmonicity on the strength of the electron-phonon coupling in aluminum hydride superconductors\cite{Rousseau2010}.
The most recent implementation of this theory was done by Ravichandran and Broido in 2018 who referred to the method as ``phonon renormalization approach"\cite{Ravichandran2018}. The motivation for his work was to include temperature effects beyond the quasiharmonic  approximation in the description of phonons and be able to describe thermal expansion and include this effect in the calculation of other properties such as lattice thermal conductivity. Although previous approaches which used the bare harmonic and anharmonic force constants were successful for many materials, they did not produce good results for simple anharmonic lattices which have large thermal expansion, such as NaCl, even at moderate temperatures.

It can be shown that the relationship between the effective force constants $\Theta$ and the zero temperature ones denoted by $\Phi$ is given by the equations defining the SCHA-4 approximation displayed in Fig, \ref{scha-diagram} and defined in eq. \ref{scha-4}. To be more precise, after substitution of the mean square displacements by their expression in terms of {\it renormalized phonon eigenvectors} $\epsilon$, the relationship, will all the indices explicitly included, becomes: 

\begin{widetext}
	\begin{equation}
		\Theta_{1,2}^{\alpha\,\beta} = \Phi_{1,2}^{\alpha\,\beta}+\frac{\hbar}{4N_q}\sum_{\vec{q}\lambda}\sum_{3,4}X_{1,2,3,4}^{\alpha\beta\gamma\delta}\frac{\epsilon_{\vec{q} \lambda}^{\tau_3\gamma}\epsilon_{-\vec{q} \lambda}^{\tau_4\delta}}{\Omega_{\vec{q}\lambda}\sqrt{m_{\tau_3}m_{\tau_4}}}e^{i\vec{q}\cdot(\vec{R}_3-\vec{R}_4)}(2n_{\vec{q}\lambda}+1)
    \label{theta}
    \end{equation}
\end{widetext}
where the numerical indices ${1,2,3,4}$ denote atom sites within the supercell, $\vec{R}_i$ are the primitive lattice translation vectors, $\vec{\tau}_i$ corresponds to atoms within the primitive cell, $N_q$ is the number of $q$ points in the Brillouin zone (equivalent to number of  unit cells in the supercell), $\Omega_{\vec{q}\,\lambda}$ is the ``renormalized" phonon frequency for mode $\lambda$ at $\vec{q}$, $\epsilon_{\vec{q} \lambda}$ is the corresponding eigenvector, and $n_{\vec{q}\lambda}$ is the Bose-Einstein occupation number. Since the renormalized phonon frequencies and eigenvectors depend in turn  on renormalized FCs $\Theta$, this equation needs to be solved self-consistently. 

In their approach, similar to the s-TDEP, atomic configurations taken from the canonical ensemble are generated in a supercell according to eq. \ref{eq:canonical}. Then forces are calculated using DFT and fitted to a 
quartic Hamiltonian to produce $\Phi$ and $X$. 
Starting with the harmonic eigenvalues and eigenvectors used in eq. \ref{theta} a first set of renormalized harmonic FCs $\Theta$ are generated. In the next iteration $\Phi$ and $X$ are kept fixed  and and $\Theta$ is updated with the new set of eigenmodes. This procedure is continued until convergent. New snapshots according to eq. \ref{eq:canonical} but with the renormalized eigenmodes and new fitting to DFT forces did not change the results significantly. 
Once the renormalized harmonic FCs are fixed, they proceed to generate a few more snapshots using the eigenmodes of the latter in  eq. \ref{eq:canonical}, and fit the DFT forces to a nonlinear force model of the form
\begin{equation}
 F_{\tau}=-\sum_{\tau'} \Theta_{\tau \tau'}U_{\tau'}-\sum_{\tau'1} \Psi'_{\tau \tau' 1} U_{\tau'} U_1 -\sum_{\tau'12} \rchi'_{\tau \tau' 12} U_{\tau'} U_1 U_2   
\end{equation}
with fixed $\Theta$. This defines the new renormalized cubic and quartic FCs $\Psi'$ and $\rchi'$, different from their zero-temperature counterparts.
 This Taylor expansion of the force is now a well-behaved form as the deviations from the harmonic part of the forces are, by construction, the smallest possible (zero on the average). This was not necessarily the case for the zero-temperature force constants. 

This renormalization framework was  applied to NaCl at temperatures of 80K and 300K, where traditional quasi-harmonic approximation(QHA) shows obvious softening effect and large lifetimes. The renormalized phonon dispersion calculated in this way matched well with inelastic neutron scattering(INS) results. It also reproduced features such as relative temperature independence of the optical phonons. Furthermore, compared to QHA, thermal expansion, phonon-phonon scattering and lattice thermal conductivity studied using renormalized phonons produced better agreements with experimental data sets. 

SCP theory once again shows its robustness for describing finite temperature phonon description and corresponding thermal properties. Similar tests were also performed using diamond to further solidify the soundness of this renormalization approach. 

%In Ravichandran's method, second-order FCs are calculated from \textit{ab initio} DFPT method using \texttt{QUANTUM ESPRESSO} package while the third and fourth-order FCs are based on same stochastic snapshot techniques recently employed by Hellman\cite{Hellman2017} and the canonical ensemble initialization formulas first proposed by Estreicher\cite{Estreicher2006} (Eq.[\ref{eq:canonical}]). Then DFT calculations are performed once again using the displacement sets in \texttt{QUANTUM ESPRESSO} to calculate forces on atoms. Later, instead of performing \textit{ab initio} MD simulation, least-square technique and translational invariance conditions combined with point-group symmetry were used to solve force-displacements equations with a truncation at quartic terms.  The scheme also takes advantage of the renormalization phonon part by 'updating' DFT forces $F_i^\alpha$ using renormalized second-order FCs $\Theta$:
%\begin{widetext}
%	\begin{equation}
%		F_i^\alpha = -\sum_{1}\Theta_{i,1}^{\alpha\,\beta}u_1^\beta-\frac{1}{2}\sum_{1,2}\Psi_{i,1,2}^{\alpha\beta\gamma}u_1^\beta u_2^\gamma-\frac{1}{6}\sum_{1,2,3}X_{1,2,3}^{\alpha\beta\gamma\delta} u_1^\beta u_2^\gamma u_3^\delta
%	\end{equation}
%\end{widetext}
%this final step ensure that renormalized FCs are still consistent with original DFT calculated dataset, thus that fitted anharmonic FCs at finite temperature are more reliable than from previous method(Eq.[\ref{eq:force}]). 

\subsubsection{Stochastic SCHA}
In 2013, Mauri's group came up with a stochastic implementation of the self-consistent harmonic approximation(SSCHA)\cite{Mauri2013-1}. This method has several noticeable features compared to previous ones. Firstly, the SSCHA equations are derived from a free energy minimization using the Bogoliubov inequality. The Hamiltonian is written as the sum of a reference effective  harmonic hamiltonian and the anharmonic corrections to it.  The variational parameters include two groups of parameters: the equilibrium position of ions $\mathbf{R}_{eq}$  and the trial harmonic FCs $\Phi_{R\tau}^{\alpha\beta}$. The minimization is performed using conjugate-gradients(CG) method. The potential energy and forces are not expanded in Taylor series but kept from DFT calculations. This has the advantage that there is no truncation and large anharmonicities can be handled exactly, but at the cost of more DFT calculations in the supercell. To alleviate the number of DFT calculations, thermal averages are computed using stochastic methods, similar to the Monte Carlo importance sampling of the canonical ensemble, as we will describe below.
To remind the reader, if the variational parameters are the equilibrium lattice positions $S_{\tau}$ and the trial force constants $\Phi$, and the free energy to be minimized is (see also eq. \ref{eq:BI}) $F_{trial}[\Phi]+\langle V[S]-V_{trial}[\Phi] \rangle$.
First a set of equilibrium atomic positions and a trial harmonic force constant $(S^0,\Phi^0)$ are guessed, and its normal modes calculated. Then many snapshots are generated from the canonical ensemble at temperature $T$ for instance from MD, or according to \ref{eq:canonical} given the normal modes of the trial FCs. These snapshots are used to compute the thermal averages of different terms needed, namely the first derivatives of the free energy which need to be set to zero in order to find the variational parameters $(S,\Phi)$.
These two minimization equations to be solved are:
\begin{eqnarray}
0&=& \frac{\partial \langle V[S] \rangle_{\Phi} }{ \partial S}  \\
0&=& \frac{\partial F_0}{ \partial \Phi} + \frac{\partial \langle V[S] \rangle_{\Phi}}{ \partial \Phi} - \frac{1}{2} \langle yy \rangle_{\Phi} \end{eqnarray}
where atomic displacements were written as $U(t) =S+y(t)$. In the second equation the derivative of the potential energy with respect to the trial FC is not zero as the averaging is done with respect to the trial density matrix $\rho_0 =e^{-\beta H_0[\Phi]}/Z_0$ involving $\Phi$ itself: $\langle V[S] \rangle= \rm Tr \rho_0[\Phi]  V[S]$
The trace is an integral over the dynamical variables $y$ and their conjugate momenta. 
Such integrals over all possible $y$ weighted by $\rho_0$ are replaced by a sum over a few atomic snapshots where $y'$s take different values. 
After the averages are calculated and the equations solved, one finds a new pair of variational parameters $(S^1,\Phi^1)$ as solution, for which the above process is repeated. This is because with the new density matrix defined by the latter averages change and the gradients are not zero anymore.
In the second and later iterations, however, new snapshots will not be generated. The same snapshots will be used, but there will be a reweighting of the probabilities according to $\rho_0[\Phi^1]/\rho_0[\Phi^0]=e^{-\beta(V_0[\Phi^1]-V_0[\Phi^0])} Z_0[\Phi^1]/Z_0[\Phi^0]$. The same reweighting will hold after iteration $l$, where $\Phi^1$ only needs to be replaced by $\Phi^l$, the FC after iteration number $l$. In these equations the subscript 0 refers to the trial harmonic quantities.

To check for the consistency of the reweighting procedure, the ratio $\frac{1}{ N_s} \sum_{j=1}^{N_s} \rho_0[\Phi^l](y_j)/\rho_0[\Phi^0](y_j) $ is checked to stay near 1. If it is very different from 1, then at that iteration, new DFT snapshots need to be generated according to the latest density matrix and trial FC, since this indicates that the original sampling was not appropriate. The question of how many snapshots are needed in order to get a decent sampling of the canonical ensemble is a subtle one and has to be found by checking the convergence of the results.

This method was later proven to be robust through various test examples such as PdH and PtH. Later the SSCHA was combined with 2n+1 theorem\cite{Mauri2015}. The scheme was capable of giving a leading-order approximation of phonon line broadening as well as possess SSCHA's ability for accurate description of phonon spectra.

The advantage of this approach is that, similar to the SCAILD and TDEP the thermal average of the ``exact" potential energy (or force) is evaluated and does not require a Taylor expansion. The evaluation of the thermal average is however stochastic as instead of an integral one is using a summation over a finite set of randomly generated snapshots.

\section{A recent extension to SCHA-4}\label{our-extension}

In this section, we will present, using the variational approach, an extension which will enable us to predict phase changes in anharmonic systems not only as a function of pressure but also the more challenging parameter of temperature. This approach is based on the self-consistent phonon theory, and therefore will use a parametrization of the potential energy as a truncated Taylor expansion. The disadvantage compared to the stochastic method is that the evaluation of the potential energy will be approximate. But the advantage is that the phase transition can be predicted with a minimal anharmonic model, and calculations are lighter. 
The extension is in the inclusion of strain and internal atomic relaxations as the temperature and unit cell shape and volume are changing.
In this case,  traditionally-dropped cubic terms become important and are explicitly present in the formulation. Coupling to other order parameters such as orbital ordering or magnetization can also be incorporated in the model.

\subsection{Formulation}
We will assume an anharmonic Hamiltonian in which the potential energy is a polynomial in atomic displacements about a reference equilibrium position.
The polynomial can be of degree 4 or more if needed. The thermal average of any power of displacements can readily be calculated.
The variational approach consists in adopting a trial but solvable Hamiltonian, which we take to be harmonic but allow displaced equilibrium positions and deformed unit cell, the optimal values of which at any given temperature can be found by minimizing the variational free energy at that temperature. 

The absolute displacement from the equilibrium site is denoted by $\vec{U}_{R\tau}(t)$. Since we want to describe phase transitions, the shape of  the unit cell may change with $T$ and as a result the internal atoms denoted by $\tau$ may find a new equilibrium position in the strained cell. For this reason,  we will write the general displacement of atoms as a sum of static strain represented by the strain tensor $\etab$,  an internal relaxation term $\bm{u}_{\tau}$,  and a dynamical displacement  $\vec{y}_{R\tau} (t)$ as follows:
\begin{gather}
	\vec{U}_{R\tau}(t) = \overline{\overline{\eta}}\cdot(\vec{R}+\vec{\tau})+\vec{u}_{\tau}+
	\vec{y}_{R\tau} (t) \label{eq:my_dis}
\end{gather}
Here the dynamical variable $\vec{y}(t)$ has a Gaussian distribution.  $R$ corresponds to the reference lattice vector, $R$ and $\tau$ uniquely define an atom in the system.   With this in mind, the typical Taylor expansion of the  potential energy (Eq.[\ref{eq:actual_pot}]) for the crystal structure can be written as:
\begin{widetext}
	\begin{equation}
	V(\vec{U}) = V_0 +\sum_{1}\Pi_1^\alpha U_1^\alpha + \frac{1}{2!}\sum_{1,2}\Phi_{1,2}^{\alpha \, \beta}\, U_{1}^{\alpha}U_{2}^{\beta}+ \frac{1}{3!}\sum_{1,2,3}\Psi_{1,2,3}^{\alpha \, \beta \, \gamma}\,U_{1}^{\alpha}U_{2}^{\beta}U_{3}^{\gamma}
	+ \frac{1}{4!}\sum_{1,2,3,4} \rchi_{1,2,3,4}^{\alpha\,\beta\,\gamma\,\delta}\,U_{1}^{\alpha}U_{2}^{\beta}U_{3}^{\gamma}U_{4}^{\delta} +\dots
	\end{equation}
	\label{eq:actual_pot}
\end{widetext}
where $\Phi,\Psi,X$ are conventional Greek letter notations for force constants of rank 2,3 and 4 respectively. The integers 1,2,3,4, which we use for brevity, refer to atoms $R_1\tau_1,R_2\tau_2,...$ in the system.  The first order FCs $\Pi$ should vanish if the expansion is done around equilibrium positions.  In order to implement Bogoliubov (Eq.[\ref{eq:BI}]) inequality for the kernel of variational approach, a harmonic trial potential model is defined with $(K,\etab,u) $ as variational parameters:
\begin{equation}
	V_{trial}(\vec{U};[K,\etab,u]) %=V_{trial}(y;[K])
	= V_0+\frac{1}{2!}\sum_{1,2}K_{1,2}^{\alpha\,\beta}y_1^\alpha y_2^\beta
	\label{eq:v_trial}
\end{equation}
here K is referred as 'trial' force constant, and dictates the distribution of $\vec{y}$. The variational parameters at a given temperature are found by minimizing the trial free with respect to them:
\begin{eqnarray}
\frac{\partial F_{trial}}{\partial K^{\alpha \beta}_{R\tau,R'\tau'}} =0  \label{var1} \\
\frac{\partial F_{trial}}{\partial \etab} =0  \label{var2} \\
\frac{\partial F_{trial}}{\partial u_{\tau}} =0  \label{var3}
\end{eqnarray}
where the trial free energy is defined as: $F_{trial}=F_0+\langle V-V_{trial} \rangle$ and thermal averages were defined in eq.\ref{trial}.
Note that in the standard SCP approach, all odd-order terms vanished 
because displacements had Gaussian distribution and only their even powers yielded non-zero thermal averages. In our formulation with strain, however, all powers of $U$ have a non-zero thermal average, and therefore cubic terms aslo explicitly appear in the results. 
Furthermore the derivatives with respect to the static variables $\etab$ and $u_{\tau}$ only appear in the potential energy $V$ but not the trial potential nor the harmonic free energy $F_0$. 

As is shown in previous variational treatments\cite{Boccara1965,Samathiyakanit1973,Glyde1976, Mauri2013-1}, the derivatives with respect to $K$ are tedious and require the use of chain rule. However the trial free energy has a more explicit dependence on $Y=\langle yy \rangle$, which has the same number of components as $K$. If we consider
\begin{equation}
	\frac{\partial \langle F_{trial}\rangle}{\partial K}=\sum\frac{\partial \langle F_{trial}\rangle}{\partial \langle y y \rangle} \frac{\partial\langle y y \rangle}{\partial K}
	\label{eq:pd_dynamic}
\end{equation}
we realize that $\frac{\partial \langle F_{trial}\rangle}{\partial K}=0$ is equivalent to $\frac{\partial \langle F_{trial}\rangle}{\partial \langle y y \rangle}=0$ as the Jacobian of the transformation ($\frac{\partial\langle y y \rangle}{\partial K}$) is not singular because physically, any change in $K$ results in a change in $Y=\langle y y \rangle$ and vice versa. 
So instead of $K$ we will minimize the trial free energy with respect to $Y$. In this case the Harmonic free energy $F_0$ does not depend on $Y$, so that the minimization equation \ref{var1} is replaced by the following equation:
\begin{equation}
    	\frac{\partial \langle V \rangle}{\partial Y} -	\frac{\partial \langle V_{trial}\rangle}{\partial Y}= \frac{\partial \langle V \rangle}{\partial Y} - \frac{1 }{2} K=0
    	\label{legendre}
\end{equation}
which gives an explicit equation for the effective harmonic force constant in terms of derivatives of the potential energy which is a polynomial in $y$ and can be readily calculated. Boccara and Sarma have shown that in general for any pair potential the derivative $\frac{\partial \langle V \rangle}{\partial Y}$ is equal to $\frac{1 }{2} \langle \frac{\partial^2  V }{\partial y \partial y }\rangle$, providing the interpretation that the optimal trial or effective harmonic FC is the thermal average of the second derivative of the potential energy with respect to atomic displacements. Their other simplifying contribution was to consider $Y$ as additional variational parameters which produced precisely eq. \ref{legendre} avoiding all the chain rules. This is in spirit similar to a Legendre transform changing the varaibles $\Phi$ to $Y=\langle yy \rangle$.

\subsection{Minimization equations with strain included}

If we denote the static part of displacement $U$ by $S$, so that $U (t) =S+y (t)$, we can easily derive the expression for the effective FC $K$. To make the equations simpler, we drop the indices and first write the thermal average of the potential energy $\langle V-V_0 \rangle$ in terms of $S$ and $Y=\langle y y \rangle$ symbolically as:
\begin{equation}
\frac{1 }{2}\Phi (Y+S^2)+\frac{1 }{6} \psi (3SY+S^3)+\frac{1 }{24} \rchi (S^4+6S^2 Y + 3 Y Y)
\end{equation}
so that the minimization equation leading to $K$ becomes
\begin{equation}
K=\Phi + \psi S +\frac{1 }{2} \rchi (S^2 + Y) 
\end{equation}
Putting back the indices, and keeping in mind that $S_1^{\alpha}= \sum_{\beta} \etab_{\alpha\beta} (R_1+\tau_1)^{\beta}+u_{\tau_1}^{\alpha} $ we obtain the expression of the effective (or renormalized) harmonic FCs  as well as the second minimization equation leading to the thermal average of the potential energy derivative, i.e. force, to be zero:
\begin{widetext}
	\begin{equation}
	K_{1,2}^{\alpha\beta}=\Phi_{1,2}^{\alpha\beta} + \sum_{3,\gamma} \psi_{1,2,3}^{\alpha\beta\gamma}\, S_3^{\gamma}+\frac{1 }{2} 
	\sum_{34,\gamma\delta} \rchi_{1,2,3,4}^{\alpha\beta\gamma\delta} \, (S_3^{\gamma}S_4^{\delta}+\langle y_3^{\gamma} y_4^{\delta} \rangle  )
	\label{K_eff}
    \end{equation}
	\begin{equation}
	0= \frac{\partial \langle V \rangle }{\partial S_1^\alpha} = \Pi_1^\alpha + \sum_{2}\Phi_{1,2}^{\alpha \, \beta}\, S_{2}^{\beta}+ \frac{1 }{2!}\sum_{2,3}\Psi_{1,2,3}^{\alpha \, \beta \, \gamma} \,(S_{2}^{\beta}S_{3}^{\gamma}+ \langle y_{2}^{\beta}y_{3}^{\gamma} \rangle)
	+ \frac{1}{3!}\sum_{2,3,4} \rchi_{1,2,3,4}^{\alpha\,\beta\,\gamma\,\delta} \,S_{2}^{\beta} (S_{3}^{\gamma}S_{4}^{\delta} +3 \langle y_{3}^{\gamma}y_{4}^{\delta} \rangle) + ...
	\end{equation}
	\label{thermal_force}
\end{widetext}
Note that once the derivatives with respect to $S$ are known, those with respect to $u$ are identical, and for $\partial/\partial \etab$ one can use the following chain rule equation:
\begin{equation}
    \frac{\partial }{ \partial \etab_{\alpha\beta}}= \frac{1 }{2} \left(\frac{\partial }{ \partial S^{\alpha}} (R+\tau)^{\beta}+
    \frac{\partial}{\partial S^{\beta}} (R+\tau)^{\alpha} \right)
\end{equation}

%There are several major advantages for this formalism: (1)computation-wise, for a d-dimensional N-atom system, the time complexity for Mauri's scheme has been reduced from $(Nd)^3$ to $(Nd)^2$ if we keep the cutoff up to fourth order terms. However Mauri's method didn't introduce this rounding error since Taylor expansion is not used. 
There are some advantages to this approach which we note below:
Since $\langle y_i y_j\rangle$ has direct correspondence with IFCs, the same point-group symmetry could be applied to significantly reduce the total number of terms. The updated trial FCs $\bm{K}$ are directly given by Eq.[\ref{K_eff}] similar to SCHA-4 and  Ravichandran's\cite{Ravichandran2018} `renormalized' phonons, but also include cubic terms coupled with other order parameters such as strain and internal relaxations. The real-space calculations are short-ranged and fast, and only the calculation of thermal averages of $\langle yy \rangle$ are done independently in reciprocal-space using the normal modes.  We use Broyden's method to solve the gradients equations in an iterative scheme since the updated $K$ will have new normal modes which define the   $\langle yy \rangle$ on the right-hand sides of eqs. \ref{K_eff} and \ref{thermal_force}.

\begin{figure}[h]
	\centering
	\includegraphics[width=\linewidth]{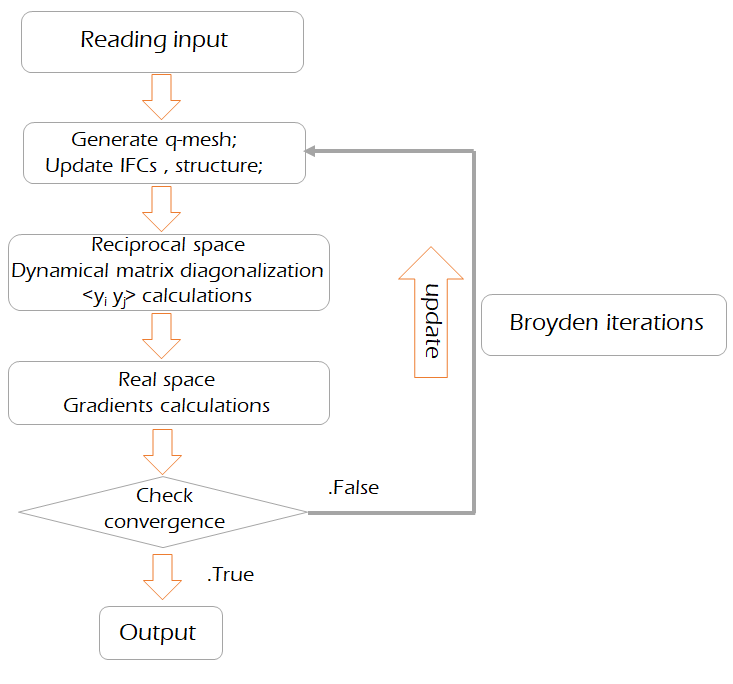}
	\caption{Computational protocol. The roots of the gradient equations are in effect found by Broyden's method, which updates the Hamiltonian parameters $S,\Phi$ in order to make the gradients zero. }
	\label{fig:protocol}
\end{figure}

The only input required besides harmonic and anharmonic FCs at finite temperature, are the structure and symmetry information. A brief flowchart for this computational protocol is displayed in Fig. \ref{fig:protocol}.

\subsection{Application to a simple model}
We applied our protocol firstly on a fabricated double-well potential in a paper studied by Morris and Gooding\cite{Morris1990}. Consider a one-dimensional atomic chain with the following potential energy $V$:
\begin{align}
	V&=V_{site}+V_{pair}  \nonumber \\
	V_{site} &= \frac{1}{2}A U_i^2 - \frac{1}{4}B U_i^4 + \frac{1}{6}C U_i^6  \nonumber \\
	V_{pair} &= \frac{k}{2}(U_i-U_{i+1})^2+\frac{k'}{4}(U_i^2+U_{i+1}^2)(U_i-U_{i+1})^2 \nonumber \\
	U_i (t) &= y_i(t) + u_0 
\end{align}
where $V_{site}$ is the on-site potential with A,B and C all positive number chosen to have a metastable minimum at $u_0=0$ and a doubly degenerate stable minima at $u_0=\pm 1$. For pair potential $V_{pair}$, the only non-zero intersite couplings are nearest-neighbor sites for the sake of simplicity. The force constants factors $k,k'$ are set to make $V_{pair}$ be much larger than the on-site potential's localization energy, to simulate the displacive limit where only weakly anharmonic oscillations happen. Notice that $U_i$ is the absolute displacement in our formalism that contains two part: (1)$u_0$ stands for internal atomic shift $\vec{u}_\tau$ (since it's mono-atomic, it's the same parameter for every i), and $y_i(t)$ is the harmonic vibration displacement. Now we could construct the trial potential energy as $V_{trial}$ which is in the standard form:
\begin{equation}
	V_{trial} = \frac{1}{2}\sum_{i,j} K_{ij}\, y_i y_j
\end{equation}
since the intersite coupling is limited to nearest neighbor and in this case it's a 1D monoatomic model, there are only 2 independent trial force constants $K_{00},K_{01}$ and so are the correlation terms $\langle y_i y_i\rangle,\langle y_i y_{i+1}\rangle $:
\begin{equation}
	\langle V_{trial}\rangle/atom = \frac{1}{2} K_{00} \langle y_i y_i\rangle + \frac{1}{2}
	K_{01} \langle y_i y_{i+1} \rangle
\end{equation} 
after taking the thermal average of $V$, we have the complete expression for the effective free energy $F_{eff}$. The goal here is optimize it with respect to three variational parameters: $K_{00},K_{01},u_0$ at every temperature. Once the convergence threshold in self-consistent iterations have been met, we use those parameters to get an approximation of real free energy value at different temperatures and by comparison, we plot the phase transition diagram in a contour plot as below:
\begin{figure}[h]
	\centering
	\includegraphics[width=\linewidth]{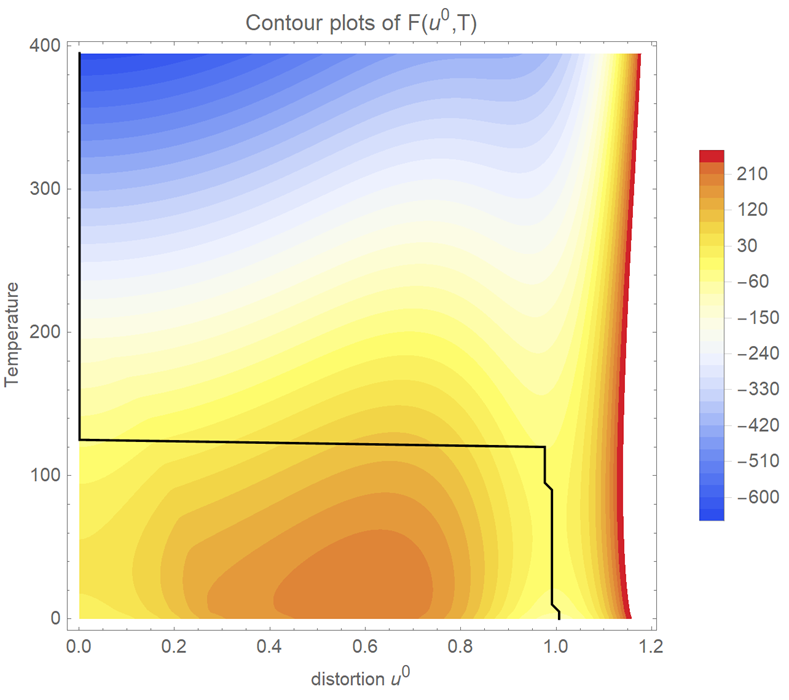}
	\caption{Contour plot of free energy $F$ with repect to atomic shift $u^0$ and temperature $T$. One can notice the phase transition from non-zero ferroic order-parameter $u^0$ at low T, to $u^0=0$ at $ T>140$. }
	\label{fig:Morris}
\end{figure}
The analytically calculated phase transition temperature that is around 140K in the papers by Morris\cite{Morris1991,Morris1995}  has been successfully reproduced using this variational approach. 

\section{Conclusions}
With the advent of accurate force and total energy calculations using DFT methods, it is now possible to implement the SCP theory to compute lattice dynamical properties of real materials at high temperatures. Several methods were described which showed the great success of this approach in predicting thermodynamic properties of a few materials at high temperatures. Few systems have so far been studied using this approach, and a lot more are awaiting the application and test of SCP to see how accurately it can predict properties of complex materials. This approach is still in its infancy and further extensions will be implemented to predict phase transitions and thermal properties of strongly anharmonic systems. 

\section{Acknowledgments}
KE thanks useful discussions with N. Ravichandran.

\appendix
\section{\\Thermodynamic properties of harmonic oscillators\label{sec:PF}}
The partition function for a single harmonic oscillator with frequency $\omega_\lambda = \epsilon_\lambda / \hbar $ is 
\begin{equation}
Z_\lambda = \sum_{n=0}^{\infty} e^{-\beta (n+1/2)\epsilon_\lambda} = \frac{e^{-\beta \epsilon_\lambda /2}}{1-e^{-\beta \epsilon_\lambda}}
\end{equation}
For each atom in the lattice, it has 3 vibrational degrees of freedom, thus for an N-atom unit cell, the partition function describes $3N$ independent oscillators:
\begin{equation}
Z_N = \prod_{\lambda=1}^{3N }\frac{e^{-\beta\epsilon_\lambda /2}}{1-e^{-\beta\epsilon_\lambda}}
\end{equation}
From which we can both acquire the Free Energy: $F = -k_B T\, {\rm ln} Z$ and phonon vibrational entropy: $S = -\frac{\partial F}{\partial T}$
\begin{align}
	F_0 &= k_B T \sum_{\lambda}^{\infty} {\rm ln[2 \,sinh} (\frac{\beta \epsilon_\lambda}{2})] \;  \label{eq:harmonicF}
	\\
	S &= k_B \sum_{\lambda}^{3N}\left( \frac{\beta \epsilon_\lambda}{2} {\rm coth}(\frac{\beta \epsilon_\lambda}{2})-{\rm ln[2\, sinh} (\frac{\beta \epsilon_\lambda}{2})]\right)
\end{align}

\section{\\Normal modes and Gaussian averages}\label{LD-EOM}
In the formulation of self-consistent phonon theory, a trial hamiltonian that has only harmonic terms is used. In SCHA theories, averages of the type $\langle yy \rangle=\rm Tr (\rho_0 \, yy)/Z_0$ need to be computed, where the density matrix itself is an exponential of a quadratic function of the variable $y$, and the trace represents a sum over all degree of freedom, which in the classical case is the integral  $\int_{-\infty}^{\infty} dy$ (see below for the quantum version).
Such Gaussian integrals are readily computed after writing the real space variables $y$ in terms of their Fourier components. 
\begin{align}
	\bm{y}_{\bm R\tau}(t)&=\frac{1}{\sqrt{N_q m_\tau}}\sum_{\bm{q}}\bm{y}_{\bm{q}\tau}(t)e^{i\bm{q}\cdot\bm{R}}\nonumber \\ 
	&=\frac{1}{\sqrt{N_q m_\tau}}\sum_{\bm{q}}e^{i\bm{q}.\bm{R}}\sum_{\lambda}y_{\bm{q}\lambda}(t)\bm{\epsilon}_{\bm{q}\lambda}^{\,\tau} \label{eq:normal} \\
\end{align}
In this way the harmonic potential energy is first decoupled and the density matrix can be written as a product over independent modes.
\begin{align}
    V&=\frac{1 }{2} \sum_{RR';\tau\tau'} K_{R\tau,R'\tau'} : \bm{y}_{\bm R\tau}\bm{y}_{\bm R'\tau'}  \nonumber \\
    &=\frac{1 }{2} \sum_{\bm q;\tau\tau'} D_{\tau\tau'}(\bm q) : \bm{y}_{\bm q\tau}\bm{y}_{-\bm q\tau'} 
\end{align}
  where $D$ is the dynamical matrix defined by:
  $$
   D_{\tau\tau'}(\bm q)=\sum_{\bm R} K_{0\tau,R\tau'}\, e^{i\bm{q}\cdot\bm{R}}
  $$
After diagonalization, its eigenvalues and eigenvectors are defined by: 
\begin{equation}
\sum_{\tau'} D_{\tau\tau'}(\bm{q})\,\cdot\,\bm{\epsilon}_{\bm{q}\lambda}^{\,\tau'}=\omega^2_{\bm{q}\lambda}\,\bm{\epsilon}_{\bm{q}\lambda}^{\,\tau} \qquad \forall \lambda \label{eq:dynamicM}
\end{equation}
This allows us, after quantization of positions and momenta, to write 
$$
	\bm y_{\bm{q}\tau}=\sum_{\lambda}  \left( \frac{\hbar}{ 2 \omega_{\bm q \lambda}} \right)^{1/2} (a_{-\bm{q}\lambda}^\dagger+a_{\bm{q}\lambda}) \,\bm{\epsilon}_{\bm{q}\lambda}^{\,\tau} 
$$
where $a^{\dagger}$ and $a$ are phonon creation and annihilation operators satisfying $\langle a^{\dagger}a \rangle = n ; \langle a a^{\dagger} \rangle = n+1 $.
Finally, from the properties of the harmonic oscillator, the orthogonality of the eigenvectors and the above results, we obtain the (quantum) thermal average of the displacements squared:
\begin{align}
\langle  \bm y_{\bm q\tau}\bm{y}_{-\bm q\tau'} \rangle &= 
\sum_{\lambda} \left( \frac{\hbar }{ 2 \omega_{\bm q \lambda}} \right) (2n_{\bm q\lambda}+1) \, \bm{\epsilon}_{\bm{q}\lambda}^{\,\tau}
\bm{\epsilon}_{-\bm{q}\lambda}^{\,\tau'} \\ 
\langle  \bm y_{\bm R\tau}\bm{y}_{\bm R'\tau'} \rangle &= 
\sum_{\bm q \lambda}  \frac{\hbar (2n_{\bm q\lambda}+1) }{ 2 N_q \omega_{\bm q \lambda}}   \, \frac{\bm{\epsilon}_{\bm{q}\lambda}^{\,\tau}
\bm{\epsilon}_{-\bm{q}\lambda}^{\,\tau'}}{\sqrt{m_{\tau} m_{\tau'}}} \, e^{i\bm q\cdot(\bm R-\bm R')} \label{eq:yy}
    \end{align}
    From the above, it can be deduced that the displacement amplitude $A_{\bm q \lambda}$ of mode $\bm q \lambda$ satisfies $A_{\bm q \lambda}^2= \frac{\hbar(2n_{\bm q\lambda}+1) }{ 2 \omega_{\bm q \lambda}} $.
One can also recover the equipartition theorem: 
\begin{equation}
	\langle \frac{p_{\bm q \lambda}^2}{2} \rangle =  \frac{1}{2}\omega_{\bm q \lambda}^2 A_{\bm q \lambda}^2 = \frac{\hbar \omega_{\bm q \lambda}}{4}\, {\rm coth} \,(\frac{\beta\hbar\omega_{\bm q \lambda}}{2})
	%\label{eq:HarmonicV}
	\label{equipart}
\end{equation}

\section{\\Formal SCHA equations\label{sec:SCP}}
If the Hamiltonian of the crystal, is:
\begin{equation}
\mathcal{H}=\sum_i-\frac{1}{2m_i} \nabla_i^2 +\sum_{i,j}V(\bm{U}_i-\bm{U}_j)
\end{equation} 
where $V$ is a pair-potential, we use a harmonic trial Hamiltonian of the form
\begin{equation}
\mathcal{H}_{trial} = \sum_i-\frac{1}{2m_i} \nabla_i^2 + \frac{1}{2}\sum_{ij} \bm{y}_i\cdot\bm{K}_{ij}\cdot \bm{y}_j
\end{equation}
where i,j are atomic labels in the lattice, and $\bm{K}_{ij}$ is the trial force constants to be determined variationally.  From the trial hamiltonian, one can write the density matrix $\rho_{trail}$ as:
\begin{equation}
\rho_{trial} = \frac{e^{-\beta\mathcal{H}_{trial}}}{\rm Tr \, e^{-\beta\mathcal{H}_{trial}}}\label{eq:rho}
\end{equation}
and the actual free energy $F$ can be approximated as:
\begin{eqnarray}
F &\approx& \rm Tr [ \rho(\mathcal{H}+\beta^{-1}\rm ln \,\rho_{trial}) ] \equiv\langle\mathcal{H}+\beta^{-1} \rm ln \, \rho_{trial}\rangle \\
&=& F_{trial}+ \langle \mathcal{H}-\mathcal{H}_{trial} \rangle
\end{eqnarray}
%while the trial free energy $F_{trial}$ is just to replace $\mathcal{H}$ by $\mathcal{H}_{trial}$ in the above expression. 
Below, we define the correlation function $\bm{Y}_{ij}$ for dynamical displacement pairs:
\begin{equation}
\bm{Y}_{ij}=\langle \bm{y}_i\,\bm{y}_j\rangle
\end{equation}
the exact expression for this term was presented in the previous section(Eq.[\ref{eq:yy}]). Now we can apply Taylor expansion around the static equilibrium positions  $\bm S_{ij}=\bm{R}_i+\bm{\tau}_i-\bm{R}_j-\bm{\tau}_j$ to calculate the thermal averaged term in actual free energy.
\begin{align}
	\langle\phi(\bm{U}_i-\bm{U}_j)\rangle &=\langle\phi(\bm{S}_{ij}+\bm{y}_i-\bm{y}_j)\rangle \nonumber \\
	&= \langle (e^{(\bm{y}_i-\bm{y}_j)}\cdot\nabla)\phi(\bm{S}_{ij})\rangle \nonumber \\
	&= e^{\frac{1}{2}\bm{Y}_{ij}:\nabla\nabla}\phi(\bm{S}_{ij})
	\label{exp2}
\end{align}
Following Boccara and Sarma\cite{Boccara1965}, the free energy $F$ can be written into a functional of both trial force constants $\bm{K}_{ij}$ and correlation functions $\bm{Y}_{ij}$, while the trial free energy $F_{trial}$ is related only to $\bm{K}_{ij}$, or rather its eigenvalues. $\bm{K}_{ij}$ and $\bm{Y}_{ij}$ can be varied independently to give two variational equations minimizing  $F$. Keeping in mind that the thermal averages depend only on $\bm K$, and $\langle \mathcal{H}_{trial} \rangle = \sum_{ij} \bm{K}_{ij}:\bm{Y}_{ij}/2$, and using eq.[\ref{exp2}], we have:
\begin{align}
	\frac{\partial F}{\partial \bm{K}_{ij}} &= \frac{\partial F_{trial}}{\partial \bm{K}_{ij}}-\frac{1}{2}\bm{Y}_{ij} =0 \label{dFdK} \\
	\frac{\partial F}{\partial \bm{Y}_{ij}} &= \frac{1}{2} \langle \nabla\nabla \phi(\bm{U}_i-\bm{U}_j)\rangle - \frac{1}{2} \bm{K}_{ij}=0  \label{k_eff}
\end{align} 
The last equation can then give an interpretation for the effective or trial force constants $\bm K$: they are the thermal average of the actual FCs over thermally displaced atoms, while eq.(\ref{dFdK})  is a general property of harmonic systems.  
Using the definition of normal modes as in eq.(\ref{eq:normal}), one can express the correlation function $\bm Y$ in terms of normal modes. This is exactly the eq.(\ref{eq:yy}). The set of equations to be solved, which constitute the SCHA approximation are then:

\begin{itemize}
    \item  assumption of normal modes (start with the harmonic ones)
    \item computation of the correlation functions $\bm Y$ from  eq.(\ref{eq:yy})
    \item  obtain the new trial FCs $\bm K$ from  eq.(\ref{k_eff}) in which use must be made of  the Taylor expansion in eq.(\ref{exp2}) and the calculated expression of $\bm Y$

    \item  calculation of normal modes of this new $\bm K$

    \item reiterate if normal modes did not converge 
\end{itemize}

%\nocite{*}

\bigskip
\bibliography{review}% Produces the bibliography via BibTeX.

\end{document}